\documentclass[11pt]{article}
\usepackage{graphicx}
\usepackage{epsfig}
\usepackage{amsmath}
\usepackage{amsfonts}
\usepackage{amssymb}
\usepackage{epstopdf}
\usepackage{amsthm}
\usepackage[usenames]{color}
\usepackage{array}
\usepackage{xcolor}
\usepackage{float}
\usepackage{subfigure}

\usepackage[
      colorlinks=true,
      linkcolor=blue,
      urlcolor=blue,
      filecolor=blue,
      citecolor=red,
      pdfstartview=FitV,
      pdftitle={},
      pdfauthor={},
      pdfsubject={},
      pdfkeywords={},
      pdfpagemode=None,
      bookmarksopen=true
]{hyperref}

\usepackage{hyperref}

\newcommand{\mysection}[1]{\section{#1}\setcounter{equation}{0}}

\def\ho{\hat\omega}
\def\wto{\widetilde\omega}

\def\be{\begin{equation}}
\def\ee{\end{equation}}
\def\beq{\begin{eqnarray}}
\def\eeq{\end{eqnarray}}

\begin{document}

\pagestyle{myheadings}
\markboth{\textsc{\small }}{%
  \textsc{\small Greybody factors for rotating black holes in higher dimensions}} \addtolength{\headsep}{4pt}


\begin{centering}
  \textbf{\Large{Greybody factors for rotating black holes in higher dimensions}}

\vspace{2cm}

  \large{Rog\'erio Jorge$^{1,\sharp}$, Ednilton S. de Oliveira$^{2,\flat}$ and Jorge V. Rocha$^{3,\natural}$}

\vspace{1cm}

\begin{minipage}{.9\textwidth}
  \small \it
    \begin{center}
$^1$
Instituto Superior T\'ecnico -- IST,
Universidade de Lisboa -- UL,\\
Avenida Rovisco Pais 1, 1049-001 Lisboa, Portugal
\\ $ \, $ \\

$^2$
Faculdade de F\'isica, Universidade Federal do Par\'a, 66075-110, Bel\'em, PA, Brazil
\\ $ \, $ \\

$^3$
Centro Multidisciplinar de Astrof\'isica -- CENTRA,
Instituto Superior T\'ecnico -- IST,\\
Universidade de Lisboa -- UL,
Avenida Rovisco Pais 1, 1049-001 Lisboa, Portugal
\\ $ \, $ 
    \end{center}
\end{minipage}

\end{centering}

\vspace{1.5cm}

\begin{abstract}
We perform a thorough study of greybody factors for minimally-coupled scalar fields propagating on the background of rotating black holes in higher (odd) dimensions with all angular momenta set equal. For this special case, the solution enjoys an enhanced symmetry, which translates into the advantageous feature of being cohomogeneity-1, i.e., these backgrounds depend on a single radial coordinate. Our analysis contemplates three distinct situations, with the cosmological constant being zero, positive or negative.
Using the technique of matched asymptotic expansions we compute analytically the greybody factors in the low-frequency regime, restricting to s-wave scattering. Our formulas generalize those obtained previously in the literature for the static and spherically symmetric case, with corrections arising from the change in the horizon area due to rotation. It is also proven that, for this family of black holes, the horizon area is a decreasing function of the spin parameter, without regard of dimensionality and of cosmological constant. Through an improvement on a calculation specific to the class of small black holes in anti-de Sitter and not restricted to the usual low-frequency regime, we uncover a rich structure of the greybody spectrum, more complex than previously reported but also enjoying a certain degree of universality.
We complement our low-frequency analytic results with numerical computations valid over a wide range of frequencies and extend them to higher angular momentum quantum numbers, $\ell>0$. This allows us to probe the superradiant regime that is observed for corotating wavefunctions. We point out that the maximum amplification factor for intermediate-size black holes in anti-de Sitter can be surprisingly large.
\end{abstract}

\vfill

\noindent
\rule{6.2cm}{0.2mm}
\\
\noindent
{\small
$^\sharp$ \texttt{rogerio.jorge@tecnico.ulisboa.pt}\\
$^\flat$ \texttt{ednilton@pq.cnpq.br}\\
$^\natural$ \texttt{jorge.v.rocha@tecnico.ulisboa.pt}
}

\thispagestyle{empty}
\newpage

\tableofcontents

\mysection{Introduction}

The thermal nature of black holes (BHs) unveiled by Hawking~\cite{Hawking:1974sw} rests on the semi-classical computation of the creation and subsequent emission of particles from the event horizon. This celebrated calculation revealed an exact blackbody spectrum. Nevertheless, fields propagating on a nontrivial gravitational background `feel' an effective potential and the emission spectrum perceived at asymptotic infinity is deformed away from a perfect blackbody by the so-called greybody factors.

The study of greybody factors --- and, in asymptotically flat spacetimes, the absorption cross-section related to them --- has a long history, dating back to the early works of Starobinskii and Churilov~\cite{Starobinskii:1973,Starobinskii:1973b}, Page~\cite{Page:1976df,Page:1976ki} and Unruh~\cite{Unruh:1976fm}. There is of course a vast literature on the subject (see~\cite{Kanti:2014vsa} and references therein) but the number of studies regarding higher dimensional BHs with non vanishing cosmological constant is limited. This very broad class of spacetimes was precisely the context in which Ref.~\cite{Harmark:2007jy} developed the computation of greybody factors for scalar fields, employing analytic methods. This was made possible by resorting to some approximations well justified in certain regimes of parameters (s-waves and low frequencies, for instance) but also by restricting to static, spherically symmetric spacetimes.
Even less work has been devoted to the more generic case in which the background is rotating --- a rare exception being Ref.~\cite{Doukas:2009cx}, which is dedicated to the study of the emission of certain graviton modes.\footnote{See also Ref.~\cite{Macedo:2013afa} for a recent analysis of the absorption of scalar fields by four-dimensional Kerr black holes.}

This scarcity of literature concerning greybody factors for rotating and non-asymptotically flat black holes is in part due to the technical complexity of the problem. In fact, the methods of~\cite{Harmark:2007jy} (see also~\cite{Creek:2006ia}) apply to BH spacetimes that depend on a single (radial) coordinate. Such geometries are also referred to as cohomogeneity-1 spacetimes, and this property simplifies the computation of black hole scattering processes by reducing it to the problem of solving a single ordinary differential equation. Although not obvious, there is a larger family of higher dimensional black holes that falls in this cohomogeneity-1 category, namely odd dimensional rotating BHs with all angular momenta equal~\cite{Kunduri:2006qa}. Consequently, the analysis of~\cite{Harmark:2007jy} is expected to generalize to these rotating backgrounds. Such an extension is the goal of this paper.\footnote{Charged and cohomogeneity-1 black hole solutions are not known explicitly and therefore we restrict to the uncharged case.}

The investigation of the emission and absorption of waves by higher dimensional black holes has experienced two surges in the past. The first period, in which mostly greybody factors for charged black holes and black branes were studied~\cite{Das:1996wn,Maldacena:1996ix,Klebanov:1997kc,Cvetic:1997uw}, was a precursor of the AdS/CFT conjecture~\cite{Aharony:1999ti}. The second revival followed the proposal of TeV gravity~\cite{ArkaniHamed:1998rs} and braneworld models~\cite{Randall:1999ee,Randall:1999vf} and the possibility of creation of BHs at particle colliders~\cite{Giddings:2001bu,Dimopoulos:2001hw}. Here the focus was on emission of waves of various spins either on the brane or in the bulk~\cite{Ida:2002ez,Ida:2005ax,Duffy:2005ns,Cornell:2005ux,Cardoso:2005mh,Creek:2007pw,Casals:2008pq,Kanti:2009sn,Sampaio:2009ra,Kanti:2010mk}.

Greybody factors carry important information about black holes and their holographically dual incarnations~\cite{Maldacena:1996ix}. Thus, it is of intrinsic interest to explore absorption probabilities for black hole geometries as general as possible, including non asymptotically flat spacetimes. While the consideration of higher dimensional, asymptotically anti-de Sitter (AdS) black holes are obviously relevant from the AdS/CFT perspective, this can also be said of Randall-Sundrum braneworld models~\cite{Randall:1999ee,Randall:1999vf}, which actually require the introduction of a negative (bulk) cosmological constant. In this context, high energy collisions in the brane with a finite impact parameter naturally produce singly rotating black holes. The analytical techniques we employ are restricted to odd dimensions and to a complementary set of solutions, with all angular momenta equal. In this sense, the present work covers a previously unaddressed portion of the parameter space of solutions to the higher dimensional cosmological Einstein equations.

There are certain advantages of the approach taken in this paper in comparison with the rest of the literature on greybody factors for rotating higher dimensional black holes. First, our computation does not rely on the Newman-Penrose formalism --- which was only recently extended to higher dimensions~\cite{Durkee:2010xq} --- to obtain a Teukolsky-like separated equation for the radial profile. Related to this, the angular eigenfunctions and their associated eigenmodes are exactly known and we do not need to compute them numerically or perturbatively in a small rotation expansion. Therefore, our results are valid for arbitrary (under-extremal) values of the rotation parameter. Secondly, and similar to Ref.~\cite{Harmark:2007jy}, we are able to treat the cases with or without a cosmological constant in a unified manner for most of the analysis, with the only differences arising in the far region asymptotics.

\subsection{Main results}

The cohomogeneity-1 setup we adopt allows for excellent control over our analytic computations of greybody factors for scalar fields in rotating backgrounds. Generically this holds in the low-frequency regime, but for the specific case of `small' black holes in AdS --- defined as BHs with horizon radius much smaller than the length scale set by the cosmological constant --- our analytic results can extend over a broad range of frequencies~\cite{Harmark:2007jy}, displaying impressive agreement with numerical calculations of the greybody factors.

Our main analytic results for the s-wave modes ($\ell=m=0$, where $\ell$ represents the total angular momentum quantum number and $m$ stands for the magnetic quantum number), which generalize the static case~\cite{Harmark:2007jy}, are the following:
\begin{itemize}
\item For asymptotically flat black holes the greybody factor for low frequencies --- denoted by $\omega$ throughout the manuscript --- behaves as $\omega^{D-2}$, where $D$ is the dimensionality of the spacetime, and the low-energy absorption cross section is equal to the horizon area, just like in the static case~\cite{Das:1996we,Harmark:2007jy}. The horizon area decreases with rotation and therefore the low-energy absorption cross section is maximised for static black holes.
\item For asymptotically de Sitter (dS) small black holes the greybody factor goes to a positive constant at low frequencies, determined by the ratio of the areas of the BH horizon and of the cosmological horizon. Therefore, the Hawking emission spectrum is enhanced at low energies, as first noted in the static case~\cite{Kanti:2005ja}. However, since the horizon area decreases with rotation, this enhancement is in part softened in the presence of angular momentum.
\item For generic asymptotically anti-de Sitter black holes the greybody factor for low frequencies behaves as $\omega^{D-2}$ and is inversely proportional to the area of the event horizon.
\item For the particular case of small AdS black holes the greybody spectrum reveals a very rich structure: superposed on a smooth increasing curve it presents large-amplitude oscillations with a periodicity consistent with the spacing between the normal modes of pure AdS spacetime. This behavior is reminiscent of the one first observed in Ref.~\cite{Harmark:2007jy} which, somewhat surprisingly, inferred the existence of critical frequencies corresponding to full transmission or full reflection, the latter occurring precisely at the normal frequencies of AdS. Here, besides generalizing to include rotation, we also improve the matching calculation and show that (i) full reflection is never achieved, nor is full transmission,\footnote{Based on the analytic expression obtained via the matched asymptotic expansion approximation, full transmission would be asymptotically obtained at high frequencies, but this is beyond the regime of validity of the computation.} (ii) and that the frequencies at which the greybody factor is suppressed appear shifted relative to the normal frequencies of AdS. As we discuss in Section~\ref{sec:smallAdS_analytic}, the spectrum displays a curious degree of universality for a wide range of frequencies, with the greybody factor depending only on the dimensionality (and possibly on the quantum numbers of the modes) up to an overall scale. The significance of this complex structure of the greybody factors for small AdS black holes is still unclear and deserves further study.
\end{itemize}

The findings mentioned above are confirmed by numerical evaluations of the greybody factor: we observe good agreement in the regimes of validity of the analytic expressions obtained. Furthermore, the numerical computations allow us to extend the results to frequency regimes outside the validity of the analytical approximations and to modes with $\ell\neq0$. In particular, this allows us to probe superradiant regimes, which are frequency intervals over which the amplitude of the reflected wave is larger than that of the incident wave. When this happens the greybody factor becomes negative.
Our main conclusions drawn from the numerical results of the greybody factor are the following:

\begin{itemize}
\item In general, we observe that the analytical results become better approximations to the exact greybody factor as the dimensionality grows.
\item As usual, the mode $\ell=m=1$ is the one that features the strongest superradiance effect, i.e., it is the mode whose minimum of the greybody factor becomes more negative. This minimum approaches zero as $D$ grows, meaning that higher spacetime dimensionalities weaken superradiance.
\item For asymptotically flat black holes the greybody factor for $\ell=m=0$ modes depends weakly on the rotation parameter.
\item For asymptotically de Sitter black holes the low-frequency analytic expressions agree well with the numerical results only for the class of {\em small} black holes, in which case the greybody factor is similar to that obtained for asymptotically flat BHs, as expected. For the {\em large} black holes (which lie outside the regime of validity of our analytic computation) the low-frequency behavior is significantly different. This class depends strongly on the rotation parameter $a$. For modes $m>0$ the superradiant frequency interval has a finite lower bound, as well as the usual upper bound. For small dS black holes the lower bound is hardly noticeable but as we increase the mass of the black hole the lower bound increases, while the upper bound decreases. For the large black holes in dS the overall greybody factor decreases as $D$ grows.
\item For asymptotically anti-de Sitter black holes we find analytic results in the low-frequency and small-black holes regimes for the s-wave, as mentioned previously. The agreement of our numerics with the latter is particularly impressive: in the small AdS black hole regime the interval over which analytic and numerical results coincide can be very broad, despite the very non-trivial behavior displayed by the greybody factor. The greybody factor depends weakly on the rotation  parameter also for this class of BHs and, as in the asymptotically de Sitter and asymptotically flat cases, superradiance occurs if $m > 0$ (and $a\neq0$). Moreover, we observe that the greybody factor at very high frequencies does not approach unity; instead, it tends to a decreasing function that is independent of the black hole mass and spin parameters.
\item The complex behavior indicated by the analytic results for the $\ell=0$ modes for small AdS black holes is still present for $\ell\neq0$, i.e., the greybody factor shows large amplitude oscillations separated by a spacing consistent with the normal mode spectrum of pure AdS. For $m>0$ this structure propagates to the superradiant frequency interval, where it becomes negative. For (near-)extremal black holes, the angular velocity of the horizon, $\Omega_h$, decreases as one increases the mass of the BH, and consequently the superradiant frequency window, $0<\omega<m\Omega_h$, shrinks. Our numerical results suggest that this negative peak in the greybody factor first increases (in absolute value) as we increase the mass of the black hole, reaches a maximum for intermediate-size black holes in AdS, and then decreases as we continue towards the class of large BHs. Furthermore, we observe surprisingly strong superradiance for the intermediate-size black holes in AdS. For both this class and the large black holes, it is challenging to obtain numerical convergence around the superradiant peak, but preliminary results indicate that the maximum amplification factor can be at the order of several hundred percent in five dimensions, for nearly maximal spin and $\ell=m=1$ modes.
\end{itemize}

\subsection{Outline}

The outline of the paper is as follows. In the next section we present the background geometries we will work with and discuss their relevant properties. Section~\ref{sec:GBF_analytic} is devoted to the analytic computation of the greybody factor in the low-frequency regime, as well as a complementary calculation that can be performed for small AdS black holes. In section~\ref{sec:GBF_numerics} we outline the numerical implementation of the computation of the greybody factors, for which we are able to obtain results also for higher partial waves. Our numerical results are presented in this section, accompanied by a comparison with our analytic findings. In the appendix we discuss the formula for the horizon area of this class of black holes and how it varies with the spin parameter.

Throughout the manuscript we employ geometrized units, in which the speed of light and the gravitational constant are both set to one, $G=c=1$.

\mysection{Odd dimensional black holes with maximally equal spins
\label{sec:ES-MP-AdS}}

We consider a massless, minimally coupled scalar field propagating on the geometry of odd dimensional, equally rotating black holes, possibly with a cosmological constant.
Thus, we will be concerned with solving the Klein-Gordon equation for this class of backgrounds. This means, in particular, that we are limiting ourselves to a regime in which the scalar field does not backreact on the geometry.

The black hole solutions we shall consider --- for generic values of the rotation --- are generalizations of the well known Myers-Perry family~\cite{Myers:1986un} to include a non-vanishing cosmological constant. They were first obtained in four dimensions by Carter~\cite{Carter:1968ks}, extended to five dimensions in Ref.~\cite{Hawking:1998kw} and have been generalized to higher dimensions in Refs.~\cite{Gibbons:2004js,Gibbons:2004uw}.
In odd spacetime dimensions, $D=2N+3$ with $N\geq1$, one can pick $N$ orthogonal planes of rotation and the general  solution is parametrized by $N$ independent angular momenta $a_i$, in addition to a mass parameter $M$.
As shown in Ref.~\cite{Kunduri:2006qa} (and previously in Ref.~\cite{Frolov:2002xf} for the five-dimensional asymptotically flat case), when the rotation parameters are set all equal, $a_i=a$, the isometry group gets enhanced and the solution becomes cohomogeneity-1.

These spacetimes depend only on a single radial coordinate, and their metric can be written as follows~\cite{Kunduri:2006qa}:
\be
 ds^2  =  - f(r)^2 dt^2 + g(r)^2 dr^2 + r^2 \widehat{g}_{ab} dx^a dx^b 
 +\, h(r)^2 \left[ d\psi + A_a dx^a - \Omega(r) dt \right]^2\,,
\label{eq:metric}
\ee
where
\beq
g(r)^2  &=&  \left( 1 + \kappa^2r^2 - \frac{2M}{r^{2N}} + \frac{2\kappa^2Ma^2}{r^{2N}} + \frac{2Ma^2}{r^{2N+2}} \right)^{-1}\,, \\
h(r)^2  &=&  r^2 \left( 1 + \frac{2Ma^2}{r^{2N+2}} \right)\,, \qquad \Omega(r) =  \frac{2Ma}{r^{2N} h(r)^2}\,, \qquad 
 f(r)  =  \frac{r}{g(r) h(r)}\,.
\eeq
In the above equations $\widehat{g}_{ab}$ represents the Fubini-Study metric on the complex projective space ${\mathbb CP}^N$ and $A=A_a dx^a$ is its K\"ahler potential.
Explicit expressions for these quantities can be constructed iteratively in $N$ (see e.g. Ref.~\cite{Dias:2010eu}).
The above form of the metric~\eqref{eq:metric} is made possible by the fact that constant $t$- and $r$-slices are topologically $(2N+1)$-spheres and these can be written as an $S^1$ bundle over ${\mathbb CP}^N$.
For the case $D=5$, this corresponds to the familiar Hopf fibration.
The coordinate $\psi$ parametrizes the $S^1$ fiber and has period $2\pi$.

\begin{figure}
\centering
\includegraphics[width=8.6cm]{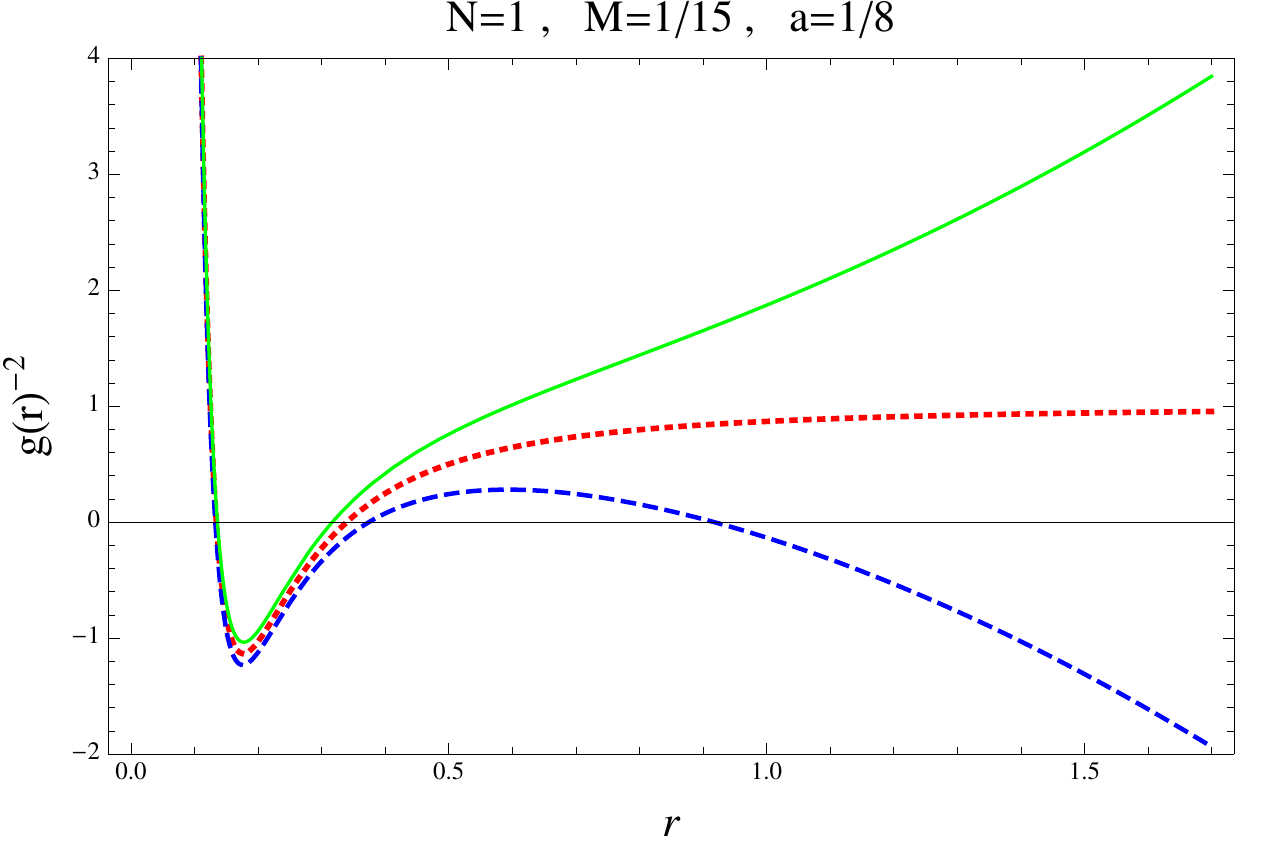}
\caption{Generic behaviour of the metric function $g(r)^{-2}$ for rotating cohomogeneity-1 black holes. The top (solid green) line represents the asymptotically AdS case, $\kappa^2=1$. The middle (dashed red) line corresponds to the asymptotically flat case, $\kappa=0$. The bottom (dashed blue) line represents the asymptotically dS case, $\kappa^2=-1$. This plot corresponds to the particular choice of parameters: $N=1, M = 1/15$ and $a=1/8$.}
\label{veff}
\end{figure}

This metric is a solution of the Einstein equations with a cosmological constant proportional to $\kappa^2$,
\be
R_{\mu\nu} = - (D-1)\kappa^2 g_{\mu\nu}\,.
\ee
The asymptotically flat case is recovered as $\kappa \rightarrow 0$.
For $\kappa^2\geq0$ the largest real root of $g^{-2}$ is denoted by  $r_h$ and marks an event horizon. For $\kappa^2<0$ the largest real root of $g^{-2}$ corresponds instead to the cosmological horizon, $r_c$, and the BH horizon is located at the second largest real root of $g^{-2}$. The behavior of $g^{-2}$ is shown in Fig.~\ref{veff}. Note that in the asymptotically AdS (dS) case it is always possible to set $\kappa^2=1$ ($\kappa^2=-1$) by appropriately rescaling the radial coordinate and the parameters.

The black hole horizon possesses the geometry of a homogeneously squashed $S^{2N+1}$. Its area is given by (see Appendix~\ref{sec:Appendix})
\be
{\cal A}_h = \Omega_{2N+1} r_h^{2N} h(r_h)\,,
\ee
where $\Omega_n\equiv 2\pi^{\frac{n+1}{2}}/\Gamma\left(\frac{n+1}{2}\right)$ is the volume of the unit $n$-sphere.
The mass ${\cal M}$ and angular momentum ${\cal J}$ of the spacetime are given by~\cite{Kunduri:2006qa}
\be
{\cal M} = \frac{\Omega_{2N+1}M}{4\pi} \left( N+\frac{1}{2} + \frac{\kappa^2a^2}{2} \right)\,, \qquad
{\cal J} = \frac{\Omega_{2N+1}Ma}{4\pi} (N+1) \,.
\label{eq:charges}
\ee
The angular velocity of the BH, which we assume to be positive without loss of generality, is defined by
\be
\Omega_h\equiv \Omega(r_h)\,,
\ee
while the black hole surface gravity can be expressed as
\be
k_h = \frac{f'(r_h)}{g(r_h)} \,,
\label{eq:surfgrav}
\ee
and is related to the Hawking temperature by $T_H= \frac{k_h}{2\pi}$.

There exists a regular extremal solution for which the horizon becomes degenerate, corresponding to a maximal, or critical, value of $a$ for given $M$ and $\kappa$.
%
In the asymptotically flat case ($\kappa=0$) we can write an explicit bound for the spin parameter, in order to guarantee the presence of an event horizon covering the curvature singularity at $r=0$:
\be
a \leq a_c \equiv \sqrt{\frac{N}{N+1}} \left(\frac{2M}{N+1}\right)^{\frac{1}{2N}} \,.
\label{eq:amax}
\ee
With a non-vanishing cosmological constant it is not possible to obtain a closed form expression for the bound on the spin parameter. The best we can do is the following. First, define dimensionless variable and parameters by
\be
\hat{r} = |\kappa| r\,, \qquad  \hat{a} = |\kappa| a\,, \qquad  \hat{M} = |\kappa|^{2N} M\,,
\ee
in terms of which the metric function $g(r)$ becomes
\be
g(\hat{r})^{-2} = 1 + \epsilon \hat{r}^2 - \frac{2\hat{M}}{\hat{r}^{2N}} + \epsilon \frac{2\hat{M}\hat{a}^2}{\hat{r}^{2N}} + \frac{2\hat{M}\hat{a}^2}{\hat{r}^{2N+2}} \,.
\ee
Here, we introduced the parameter $\epsilon$ which takes the values $-1$ and $+1$ for asymptotically de Sitter and anti-de Sitter spacetimes, respectively. Now we can give an upper bound on the dimensionless spin in terms of the black hole event horizon,
\be
\hat{a} \leq \hat{a}_c \equiv \hat{r}_h  \sqrt{\frac{N + \epsilon (N+1)\hat{r}_h^2}{(N+1)(\hat{r}_h^2+\epsilon)^2}} \,,
\label{eq:amax_with_CC}
\ee
while the horizon radius is implicitly determined as a function of the mass parameter,
\be
(\hat{r}_h^2+\epsilon)^2 \hat{r}_h^{2N} = \frac{2\hat{M}}{N+1} \,.
\label{eq:rh_with_CC}
\ee
One recovers the asymptotically flat result~\eqref{eq:amax} by taking the limit $\kappa\to0$ in Eqs.~\eqref{eq:amax_with_CC} and~\eqref{eq:rh_with_CC}, with either $\epsilon=\pm1$.\footnote{We note that explicit formulas can be found for lower dimensional cases, namely $N=1,2,4$ and $6$, for which the polynomial equation~\eqref{eq:rh_with_CC} has known solutions. At any rate the resulting expressions are not illuminating so we do not present them.}

\mysection{Analytical computation of greybody factors
\label{sec:GBF_analytic}}

The scalar wave equation we are interested in solving is the Klein-Gordon equation,
\be
\nabla^2\Phi - \mu^2 \Phi=0\,.
\label{eq:KleinGordon}
\ee
For now we keep the mass term $\mu^2$ in our equations in interest of generality, but later we will set it to zero.
Gravitational perturbations of equal angular momenta Myers-Perry black holes in odd dimensions have been analyzed in~\cite{Kunduri:2006qa,Dias:2010eu,Dias:2014eua}, where the focus was on the study of the quasinormal modes. Here instead we are concerned with (scalar) wave scattering, which amounts to a different choice of boundary conditions.

The wave equation separates under the following ansatz (see reference~\cite{Kunduri:2006qa}):
\be
\Phi = e^{-i\omega t} \Phi_\omega(r,\psi,x^a) \equiv e^{-i\omega t -i m \psi} R(r) \mathbb{Y}(x^a)\,, \qquad \qquad m \in \mathbb{Z}\,.
\ee
Here, $\mathbb{Y}$ represents an eigenfunction of the charged scalar Laplacian on $\mathbb{C}P^N$ and its eigenvalues are given by
\be
-{\cal D}^2 \mathbb{Y} = -\widehat{g}^{ab} (D_a-imA_a)(D_b-imA_b)\mathbb{Y} = \left[ \ell(\ell+2N)-m^2 \right]\mathbb{Y} \,,
\ee
where
\be
\ell = 2k + |m| \,, \qquad \qquad k=0,1,2,\dots
\ee
Note that, for a given $\ell$, the spacing between values of $m$ is always even. The gauge-covariant derivative on $\mathbb{C}P^N$ is denoted by $D_a$.

The scalar wave equation~\eqref{eq:KleinGordon} can now be cast into Regge-Wheeler form by defining
\be
- \frac{f}{g} \frac{d}{dr} \left( \frac{f}{g} \frac{d\Psi}{dr} \right) + \left[ V - (\omega-m\Omega)^2\right] \Psi = 0 \,,
\label{eq:scalarwave}
\ee
where the potential is given by
\be
V(r) = \frac{1}{\sqrt{h}\,r^N} \frac{f}{g} \frac{d}{dr} \left[ \frac{f}{g} \frac{d}{dr} \left( \sqrt{h}\,r^N \right) \right]  + \frac{f^2}{r^2} \left[ \mu^2 r^2 + \ell(\ell+2N) - m^2\left( 1-\frac{r^2}{h^2} \right) \right] \,,
\label{eq:potential}
\ee
and
\be
\Psi(r) = \sqrt{h(r)}\,r^N R(r)\,.
\ee
In terms of the tortoise coordinate $x$, defined by $dx/dr= g/f$, equation~\eqref{eq:scalarwave} simplifies to
\be
- \frac{d^2\Psi}{dx^2} + \left[ V - (\omega-m\Omega)^2\right] \Psi = 0 \,.
\label{eq:wavetortoise}
\ee

In this section we will obtain expressions in closed form for the greybody factors corresponding to low-frequency absorption of massless scalar waves in three distinct cases: $\kappa^2=0$, $\kappa^2<0$ and $\kappa^2>0$, corresponding to asymptotically flat, dS and AdS spacetimes, respectively.  We will follow the approach of~\cite{Harmark:2007jy}, which was developed to deal with static, spherically symmetric and higher dimensional black holes. As we will see, this framework can be extended to the case of rotating, cohomogeneity-1 BH spacetimes and, similarly, it requires us to restrict to the massless case, $\mu=0$, and moreover to `s-wave' modes, $\ell=m=0$ (i.e., the lowest harmonic on the squashed $(2N+1)$-sphere).\footnote{Ref.~\cite{Maldacena:1997ih}, working in four dimensions, only required $\omega M\ll 1$ for the low-frequency matching procedure and was able to obtain closed expressions also for modes with $\ell\neq0$. However, such an approach does not succeed here: the near-horizon region wave equation that one obtains in $D=2N+1>4$ is not of the hypergeometric form and thus cannot be solved exactly, as is the case in $D=4$.} Thus, for the rest of this section we will consider $\mu=\ell=m=0$, unless noted otherwise.

The computation proceeds by splitting the domain of outer communications into three (possibly overlapping) regions --- where approximate solutions can be found analytically --- and then gluing these solutions. In order to be able to match the wavefunctions across different regions the procedure thus requires the wavelength to be large compared to the typical length scales in the BH geometry.
In the asymptotically flat case the only available scale is the BH temperature\footnote{To be precise, the parameter $a$ also defines a length scale in addition to the mass $M$. However, this extra scale cannot be arbitrarily large because it is limited by the extremality bound, $a^2\leq a_c^2 \equiv \frac{N}{N+1}r_h^2$.} and the low-frequency regime is just defined by
\be
\omega \ll T_H\,.
\label{eq:lowfreqAF}
\ee
In the asymptotically (A)dS case the cosmological constant introduces another length scale and other restrictions might be necessary for validity of the analytical results, e.g. the limitation to small black holes (compared to the length scale set by the cosmological constant, $\kappa^{-1}$) in Sections~\ref{sec:dS_analytic} and~\ref{sec:smallAdS_analytic}.

The approximate solutions in the near horizon region and intermediary region do not depend on the asymptotics of the spacetime. Consequently we will treat the different cases $\kappa^2=0, \kappa^2<0, \kappa^2>0$ in a unified manner for these two regions. On the contrary, the far region is affected by the cosmological constant so this region must be treated separately for different asymptotics.

\subsection*{The near-horizon region (region I)}

The near-horizon region is defined by
\be
r \simeq r_h \,, \qquad V(r) \ll \omega^2\,.
\label{eq:region1}
\ee
Performing a series expansion near the horizon and using $\partial_r \left[g(r)^{-2}\right]_{r_h} = 2h(r_h)k_h/r_h$ we have 
\beq
\frac{f(r)}{g(r)} &=& \frac{r}{g(r)^2 h(r)} = 2k_h (r-r_h) + O(r-r_h)^2\,, \nonumber \\
f(r)^2 &=& 2k_h \frac{r_h}{h(r_h)} (r-r_h) + O(r-r_h)^2\,.
\eeq
Thus, the potential in this region is approximated by
\be
V(r) \simeq 2 k_h^2 \left[N + (N+1)\frac{r_h^2}{h(r_h)^2}\right] \frac{r-r_h}{r_h}\,,
\ee
and condition~\eqref{eq:region1} together with~\eqref{eq:lowfreqAF}, translates into
\be
\frac{r-r_h}{r_h} \ll \frac{1}{2\left[N + (N+1)\frac{r_h^2}{h(r_h)^2}\right]} \frac{\omega^2}{k_h^2} \ll 1\,.
\ee

Now, in the regime~\eqref{eq:region1} the wave equation~\eqref{eq:wavetortoise} becomes
\be
\frac{d^2\Psi}{dx^2} + \omega^2 \Psi = 0 \,,
\ee
with solution
\be
\Psi = B^{+} e^{-i\omega x}  \qquad\qquad \Longrightarrow \qquad\qquad  \Phi_\omega \sim A_I e^{-i\omega x} \qquad\text{for} \;\; r\simeq r_h \,.
\ee
Here, $B^{+}$ is a constant. Generically the quantity $A_I$ is a function of the coordinates $\psi$ and $x^a$, but in the case $\ell=m=0$ we are considering, it is also a constant, proportional to $B^{+}$. We have picked the solution that is purely ingoing at the horizon, which is the physically relevant boundary condition if one wants to consider the process of scattering of a wave coming in from infinity.

The flux at the horizon can then be computed as
\be
J_{hor} = -\frac{{\cal A}_h}{2i} \left( \Phi_\omega^* \frac{d\Phi_\omega}{dx} - \Phi_\omega \frac{d\Phi_\omega^*}{dx} \right) 
= \Omega_{2N+1} r_h^{2N} h(r_h)\, \omega |A_I|^2 \,.
\label{eq:Jhor}
\ee

For $r\simeq r_h$ the tortoise coordinate is $x \simeq \frac{1}{2k_h} \log\left(\frac{r-r_h}{r_h}\right)$. 
So, slightly away from the horizon the solution behaves like
\be
\Phi_\omega(r) \simeq A_I \left[ 1 - i\frac{\omega}{2k_h} \log\left(\frac{r-r_h}{r_h}\right) \right] \,.
\label{eq:region1expansion}
\ee
%

\subsection*{The intermediate region (region II)}

The intermediate region is defined by
\be
V(r) \gg \omega^2\,.
\ee
In this regime the wave equation~\eqref{eq:scalarwave} can be written as
\be
- \frac{1}{\sqrt{h}\,r^N} \frac{f}{g} \frac{d}{dr} \left[ \left(\sqrt{h}\,r^N\right)^2 \frac{f}{g} \frac{d}{dr}\left( \frac{\Psi}{\sqrt{h}\,r^N} \right) \right]  = 0 \,,
\ee
and the general solution is
\be
\Phi_\omega(r) = \frac{\Psi(r)}{\sqrt{h(r)}\,r^N} = A_{II} + B_{II} \int^r_\infty \frac{g(r')^2}{r'^{2N+1}}dr'  \,.
\label{eq:region2}
\ee

Near the horizon, $r\simeq r_h$, this solution is approximated by
\be
\Phi_\omega(r) \simeq A_{II} + \frac{B_{II}}{2k_h h(r_h) r_h^{2N}} \log\left(\frac{r-r_h}{r_h}\right) \,.
\ee
This allows us to match the solution in this intermediate region with the solution found for the near-horizon region.
Comparing with equation~\eqref{eq:region1expansion} we obtain
\be
A_{II} = A_I\,, \qquad\qquad  B_{II} = -i\omega\, h(r_h)\, r_h^{2N} A_I \,.
\ee

Now let us investigate the behavior of~\eqref{eq:region2} as we approach the far region, $r\gg r_h$, where the metric function $g(r)^2$ is well approximated by $(1+\kappa^2 r^2)^{-1}$. Here we will have to consider separately the different asymptotics:

\begin{itemize}
\item In the asymptotically flat case the solution~\eqref{eq:region2}, with $A_{II}$ and $B_{II}$ already fixed in terms of $A_I$, becomes
\be
\Phi_\omega(r) \simeq A_I \left[ 1 + i\omega\frac{h(r_h)\, r_h^{2N}}{2N r^{2N}} \right] \,.
\label{eq:region2expansionAF}
\ee
\item In the asymptotically dS case the result is the same as~\eqref{eq:region2expansionAF} if we consider going to large $r$ (compared with the BH horizon radius, $r_h$) but still in the region defined by $r \ll |\kappa|^{-1}$. For this to be possible we need to restrict to `small'  de Sitter black holes, $|\kappa|r_h\ll 1$, in which case the cosmological horizon occurs at $r_c \simeq |\kappa|^{-1}$. 
\item In the asymptotically AdS case, going to a sufficiently distant zone, $\kappa r\gg 1$, the solution~\eqref{eq:region2} becomes
\be
\Phi_\omega(r) \simeq A_I \left[ 1 + i\omega\frac{h(r_h)\, r_h^{2N}}{2(N+1)\kappa^2 r^{2N+2}} \right] \,.
\label{eq:region2expansionAdS}
\ee
\end{itemize}
Expressions~\eqref{eq:region2expansionAF} and~\eqref{eq:region2expansionAdS} will be used below to match to the far region solutions.

\subsection{Asymptotically flat case}

\subsubsection*{The asymptotic region (region III)}

The far region is defined by $r\gg r_h$, in which case the various metric functions are well approximated by
\be
g(r)\simeq1\,, \qquad  f(r)\simeq1\,, \qquad  h(r)\simeq r\,, \qquad  \Omega(r)\simeq0\,.
\ee
Introducing $\rho\equiv \omega r$, the scalar wave equation~\eqref{eq:scalarwave} in this far region (keeping factors of $\ell$ and $\mu$) becomes
\be
\frac{d^2 \Phi_\omega}{d\rho^2} + \frac{2N+1}{\rho}\frac{d \Phi_\omega}{d\rho} + \left[ 1-\frac{\mu^2}{\omega^2} - \frac{\ell(\ell+2N)}{\rho^2} \right] \Phi_\omega \simeq 0 \,,
\ee
whose solution is given in terms of Hankel functions:
\be
\Phi_\omega(\rho) = \rho^{-N} \left[ C_1 H^{(1)}_{N+\ell}(\sqrt{1-\mu^2/\omega^2}\rho) + C_2 H^{(2)}_{N+\ell}(\sqrt{1-\mu^2/\omega^2}\rho) \right] \,.
\label{eq:region3hankel}
\ee
For the matching calculation we will be interested in setting $\mu=\ell=0$ in the above expression, which then reduces to eq.~(2.32) of Ref.~\cite{Harmark:2007jy}. Keeping only the leading terms in powers of $1/r$ yields
\be
\Phi_\omega \simeq \frac{C_1+C_2}{2^{N} \Gamma(N+1)}+i(C_2-C_1)\frac{2^N \Gamma(N)}{\pi \omega^{2N}r^{2N}}\,.
\label{eq:region3expansion}
\ee

Matching the coefficients of \eqref{eq:region2expansionAF} and \eqref{eq:region3expansion}, we get
\be
C_1+C_2= A_I 2^N \Gamma(N+1), \quad C_2-C_1=A_I \frac{\pi \omega^{2N+1}r_h^{2N} h(r_h)}{2^{N+1} \Gamma(N+1)}\,.
\label{eq:matchingCA}
\ee
%

\subsubsection*{Greybody factor and absorption cross section}

We are now in position to compute the greybody factor.
It is given by the ratio of the flux at the horizon, $J_{hor}$, by the ``incoming'' flux at spatial infinity, $J_{in}$.
The flux at spatial infinity is obtained through
\be
J_{asy} = -\lim_{r\to\infty}\Omega_{2N+1} r^{2N+1} \frac{1}{2i} \left( \Phi_\omega^* \frac{d\Phi_\omega}{dr} - \Phi_\omega \frac{d\Phi_\omega^*}{dr} \right) 
=  \frac{2\Omega_{2N+1} \omega^{-2N}}{\pi} \left( |C_2|^2 - |C_1|^2 \right) 
\equiv J_{in} - J_{out} \,.
\label{eq:JasyAF}
\ee
It can be verified using~\eqref{eq:matchingCA} that $J_{asy}=J_{hor}$, which is a simple consequence of the total flux being conserved. The greybody factor is then given by
\be
\gamma(\omega) =\frac{J_{hor}}{J_{in}} = 1-\frac{|C_1|^2}{|C_2|^2}\simeq 4 \frac{C_2-C_1}{C_1+C_2} = \frac{4 \pi\, \omega^{2N+1}r_h^{2N+1}}{2^{2N+1} [\Gamma(N+1)]^2}\frac{h(r_h)}{r_h}\,.
\label{eq:af-low_freq_gbf}
\ee
The approximation performed in the third equality is valid for $\omega r_h\ll 1$, which then supplements the condition~\eqref{eq:lowfreqAF} in defining the regime of validity of the matching calculation for the asymptotically flat black holes we consider. However, for asymptotically flat BHs $T_H \simeq 1/r_h$ and so this condition is equivalent to~\eqref{eq:lowfreqAF}.

To obtain the low-frequency absorption cross section one must simply multiply the greybody factor by a coefficient that projects a plane wave onto an ingoing spherical wave~\cite{Gubser:1997qr}~\footnote{This result is based on the optical theorem for spherically symmetric scattering potentials. However, the spacetimes we study display spherical symmetry asymptotically, so its use is justified.}
\be
\sigma(\omega)=\frac{(2\pi)^{2N+1}}{\omega^{2N+1}\Omega_{2N+1}}\gamma(\omega) \;\;
\stackrel{\omega\to0}{\longrightarrow} \;\;  \Omega_{2N+1} r_h^{2N} h(r_h) = {\cal A}_h\,.
\ee
We conclude that the low-energy absorption cross section for massless scalar waves is simply given by the horizon area. Therefore, the universal result of Ref.~\cite{Das:1996we}, valid for spherically symmetric BHs, still applies to this class of rotating black holes. This outcome might have been expected based on a theorem due to Higuchi~\cite{Higuchi:2001si}, which generalizes the result of~\cite{Das:1996we} from the static, spherically symmetric case to stationary, circularly symmetric spacetimes.\footnote{The family of black holes we are studying fall within the class of solutions considered in Ref.~\cite{Higuchi:2001si}. However, it is not clear if one of the conditions of the theorem -- concerning the convergence of asymptotically-plane-wave solutions to the scalar field equation in the zero-frequency limit -- is satisfied. Therefore, without an analysis of this technical point, we cannot rely on the result of Ref.~\cite{Higuchi:2001si}.}

As shown in Appendix~\ref{sec:Appendix}, the horizon radius $r_h$ decreases when the spin parameter $a$ is increased and $M$ is kept fixed. Consequently, the horizon area actually {\em decreases} as the angular momentum is increased, at constant physical mass. Therefore, the addition of rotation reduces the low-energy absorption cross section for these equal angular momenta black holes.

\subsection{Asymptotically dS case
\label{sec:dS_analytic}}

In the asymptotically dS case there is one extra lengthscale, namely $|\kappa|$ (recall that $\kappa^2$ is negative in  asymptotically dS spacetimes), and for the matching procedure we shall need to supplement the low-frequency condition~\eqref{eq:lowfreqAF} with the additional relation $\omega \ll |\kappa|$. Moreover, we shall need to restrict to the class of small black holes, for which $|\kappa|r_h \ll 1$. Putting everything together, the validity of the analytical expression we will obtain for the greybody factor in dS is guaranteed if
\be
\omega \ll T_H\,, \qquad
\omega \ll |\kappa|\,, \qquad
|\kappa|r_h \ll 1\,.
\label{eq:lowfreqdS}
\ee
%

\subsubsection*{The asymptotic region (region III)}

The far region is defined by $r\gg r_h$, in which case the various metric functions are well approximated by
\be
g(r)^{-2}\simeq1-|\kappa|^2r^2\,, \qquad  f(r)^2\simeq1-|\kappa|^2r^2\,, \qquad  h(r)\simeq r\,, \qquad  \Omega(r)\simeq0\,.
\ee
In this region the tortoise coordinate is $x \simeq |\kappa|^{-1} \arctan(|\kappa| r)$ and the potential~\eqref{eq:potential} becomes (here we reinstate factors of $\ell$ and $\mu$)
\be
V(r)\simeq (1-|\kappa|^2r^2)\frac{(4N^2-1)(1-|\kappa|^2r^2)-4(2N+1)|\kappa|^2r^2 + 4\ell(\ell+2N) + 4\mu^2r^2}{4r^2} \,,
\ee
which agrees with  eq.~(3.5) of Ref.~\cite{Harmark:2007jy}.
This manifestly shows that, for $\ell=\mu=0$, the region $r_h \ll r \ll r_c\simeq |\kappa|^{-1}$ is included in the intermediate region (where $V(r)\gg\omega^2$) as long as we are in the low-frequency regime as determined by Eq.~\eqref{eq:lowfreqdS}, namely $\omega \ll |\kappa|$.

Introducing $y\equiv |\kappa|^2 r^2$, the scalar wave equation~\eqref{eq:scalarwave} in this far region (still keeping factors of $\ell$ and $\mu$) becomes
\begin{flalign}
&4y(1-y)^2\frac{d^2\Psi}{dy^2} + 2(1-y)(1-3y) \frac{d\Psi}{dy}  \nonumber\\
&+\left[ \frac{\omega^2}{|\kappa|^2} - (1-y)\left(\frac{\mu^2}{|\kappa|^2} + \frac{\ell(\ell+2N)}{y}\right) - \frac{1-y}{4y} \left\{ (4N^2-1)(1-y)-4(2N+1)y \right\} \right] \Psi \simeq 0 \,,
\end{flalign}
whose solution can be given in terms of hypergeometric functions. The expression for $\mu\neq0$ and $\ell\neq0$ is exceedingly long so we avoid writing it explicitly. For the matching calculation we will be interested in setting $\mu=\ell=0$, in which case we obtain
%
%
\beq
\Psi &=& \widetilde{C}_1 \, y^{\frac{2N+1}{4}}(1-y)^{-i \frac{\ho}{2}} {}_2F_1\left( -\frac{i \ho}{2}, N+1-\frac{i\ho}{2}, 1-i\ho;\; 1-y \right) \nonumber\\
&& +\, \widetilde{C}_2 \, y^{\frac{2N+1}{4}}(1-y)^{i \frac{\ho}{2}} {}_2F_1\left( \frac{i \ho}{2}, N+1+\frac{i\ho}{2}, 1+i\ho;\; 1-y \right) \,.
\label{eq:solutionDS}
\eeq
where $\ho\equiv \omega/|\kappa|$.\footnote{This form of the solution is different from the one used in Ref.~\cite{Harmark:2007jy} (see Eq.~(3.9)). Although they are equivalent, the form used in Ref.~\cite{Harmark:2007jy} is not well defined in this case, since one of the hypergeometric functions has a pole coming from the third entry being a non-positive integer. The calculation performed in Ref.~\cite{Harmark:2007jy} can be regarded as an analytic continuation in the number of dimensions.}

The solution as given above is well suited to analyze the behavior near the cosmological horizon, $y\simeq1$, where it behaves as follows:
\be
\Psi \simeq \widetilde{C}_1 \, 2^{-i\ho} e^{i\omega x} + \widetilde{C}_2 \, 2^{i\ho} e^{-i\omega x} \,.
\ee
Here we have employed $y=|\kappa|^2r^2 = [\tanh(|\kappa| x)]^2$. For $\Phi_\omega$ this translates into
\be
\Phi_\omega(x) \simeq \widetilde{C}_1 \, 2^{-i\ho} |\kappa|^{N+1/2} e^{i\omega x} + \widetilde{C}_2 \, 2^{i\ho} |\kappa|^{N+1/2} e^{-i\omega x} \,.
\ee

On the other hand, using a standard identity of hypergeometric functions (see Section 15.3 of Ref.~\cite{Abramowitz:1970as}) it can be shown that the behavior for $y\simeq0$ of the solution~\eqref{eq:solutionDS} is
\be
\Psi \simeq C'_1 \, y^{\frac{2N+1}{4}} + C'_2 \, y^{-\frac{2N-1}{4}} \,.
\label{eq:solutionDS2}
\ee
The coefficients $C'_{1,2}$ introduced above are related with $\widetilde{C}_{1,2}$ through
\be
\left(
\begin{array}{c}
\widetilde{C}_1  \\
\widetilde{C}_2
\end{array}
\right)
=
\left(
\begin{array}{cc}
B_{11} & B_{12}  \\
B_{21} & B_{22}
\end{array}
\right) 
\left(
\begin{array}{c}
C'_1  \\
C'_2
\end{array}
\right)\,,
\ee
with
\beq
B_{11}(\ho) &=& \frac{(-1)^N \Gamma(N+1)\, \Gamma(1-\frac{i\ho}{2})\, \Gamma(-N-\frac{i\ho}{2})}{\Gamma(1-i\ho)\, [U(\ho)-U(\ho)^*]} = B_{21}(\ho)^*\,, \label{eq:B11}\\
B_{12}(\ho) &=& \frac{2\pi i(-1)^N \Gamma(1-\frac{i\ho}{2})\, U(\ho)}{\ho\, \Gamma(N)\, \Gamma(1-i\ho)\, \Gamma(-N+\frac{i\ho}{2}) \sin\left(\frac{i\ho\pi}{2}\right) [U(\ho)-U(\ho)^*]} = B_{22}(\ho)^*\,.
\eeq
For convenience we defined the quantity
\be
U(\ho) \equiv -\psi(1) -\psi(N+1) +\psi\left(\frac{i\ho}{2}\right) +\psi\left(N+1+\frac{i\ho}{2}\right) \,, 
\ee
where $\psi(x)$ represents the digamma function. Observe also that the $B_{ij}$ coefficients obey the following identity:
\be
B_{11}(\ho)B_{22}(\ho) - B_{12}(\ho)B_{21}(\ho) = \frac{i\pi N \sin\left(i\pi\ho\right)}{\ho\left[\sin\left(\frac{i\pi\ho}{2}\right)\right]^2 [U(\ho)-U(\ho)^*]} \stackrel{\omega\to0}{\longrightarrow} - \frac{iN}{\ho} \,.
\label{eq:Bidentity}
\ee

The solution~\eqref{eq:solutionDS2} expressed in terms of $\Phi_\omega$, after reverting to the variable $r$, is
\be
\Phi_\omega(r) \simeq C'_1 \, |\kappa|^{N+1/2} + C'_2 \frac{|\kappa|^{1/2-N}}{r^{2N}} \,,
\ee
which upon matching with the solution~\eqref{eq:region2expansionAF} obtained in the intermediary region yields
\be
C'_1 =  |\kappa|^{-N-1/2} A_I \,, \qquad\qquad 
C'_2 = \frac{i \omega |\kappa|^{N-1/2} r_h^{2N} h(r_h)}{2N} A_I \,.
\label{eq:relateCprime}
\ee
%

\subsubsection*{Greybody factor}

Since the solution near the cosmological horizon takes the same form as the solution near the black hole horizon, namely a plane wave in the tortoise coordinate, the asymptotic flux is obtained similarly,
\be
\widetilde{J}_{asy} = -\frac{{\cal A}_c}{2i} \left.\left( \Phi_\omega^* \frac{d\Phi_\omega}{dx} - \Phi_\omega \frac{d\Phi_\omega^*}{dx} \right) \right|_{r=r_c}
= \Omega_{2N+1} r_c^{2N}h(r_c) |\kappa|^{2N+1}  \omega\, (|\widetilde{C}_2|^2 - |\widetilde{C}_1|^2) \,,
\ee
and it can be verified that for the class of small dS black holes to which our calculation is restricted --- so that $r_h \ll r_c$, implying $h(r_c)\simeq r_c$ and $|\kappa| r_c \simeq 1$ --- the asymptotic flux equals the flux at the horizon. The greybody factor is then
\be
\gamma(\omega) = \frac{J_{hor}}{\widetilde{J}_{in}} = 1- \frac{|\widetilde{C}_1|^2}{|\widetilde{C}_2|^2} 
= 1- \left| \frac{B_{11}(\ho) C'_1+B_{12}(\ho) C'_2}{B_{21}(\ho) C'_1+B_{22}(\ho) C'_2} \right|^2 \,.
\label{eq:greybodyDS}
\ee
For small dS black holes and in the low-frequency regime defined by~\eqref{eq:lowfreqdS} we have $C'_2 \ll C'_1$ and so  we may approximate the above by
\be
\gamma(\omega) \simeq 1- \left| 1+ \left(\frac{B_{12}(\ho)}{B_{11}(\ho)}-\frac{B_{22}(\ho)}{B_{21}(\ho)} \right) \frac{C'_2}{C'_1} \right|^2 \,.
\ee
Finally using~\eqref{eq:B11}, \eqref{eq:Bidentity} and~\eqref{eq:relateCprime} we obtain
\be
\gamma(\omega) \simeq \frac{(\kappa r_h)^{2N+1}}{|B_{11}(\ho)|^2} \frac{h(r_h)}{r_h} \simeq \frac{1}{|B_{11}(\ho)|^2} \frac{{\cal A}_h}{{\cal A}_c} \,.
\label{eq:GBF_dS}
\ee
In particular, in the zero-frequency limit $|B_{11}(\ho)|^2\to 1/4$ and so the greybody factor goes to a finite constant which is proportional to the ratio of the black hole and cosmological horizons also in the presence of a non-zero rotation parameter, as in the non rotating case studied previously~\cite{Brady:1996za,Kanti:2005ja,Harmark:2007jy}. Since the horizon area decreases with rotation (see Appendix~\ref{sec:Appendix} for details) the inclusion of spin reduces the low-energy limit of the greybody factor, at least for small de Sitter black holes for which the effects of rotation on the cosmological horizon are negligible. Nevertheless, we shall see in Sec.~\ref{sec:GBF_numerics} that this conclusion also applies to large dS black holes.

Greybody factors for non-minimally coupled --- or, equivalently, massive --- scalar fields in four-dimensional Schwarzschild-de Sitter were analyzed in Ref.~\cite{Crispino:2013pya} (and very recently in~\cite{Kanti:2014dxa}) where it was shown that this peculiarity of the low-frequency greybody factor approaching a finite constant is absent as soon as the scalar becomes massive. We expect a similar behavior in higher dimensions and for rotating black holes.

\subsection{Asymptotically AdS case}

In the asymptotically AdS case there is also one extra length scale, namely $\kappa$, and to proceed with the matching program we shall need to supplement the low-frequency condition~\eqref{eq:lowfreqAF} with the additional relation
\be
\omega \ll \kappa\,.
\label{eq:lowfreqAdS}
\ee
%

\subsubsection*{The asymptotic region (region III)}

Once again, the asymptotic region is defined by $r \gg r_h$, so that the potential obeys (here we reinstate factors of $\ell$ and $\mu$)
\be
V(r) \simeq \left( 1+\kappa^2r^2 \right) \left[ \frac{N^2-1/4 + \ell(\ell+2N)}{r^2} + \mu^2 + \kappa^2\left(N+\frac{1}{2}\right)\left(N+\frac{3}{2}\right) \right] \geq \left( N^2-\frac{1}{4} \right) \kappa^2 \gg \omega^2 \,.
\ee
In the last step we used the low-frequency condition~\eqref{eq:lowfreqAdS}.
Thus, this region is included in the intermediate region and we can match the solutions for the two regions.

In particular, for $\kappa r\gg 1$ the wave equation becomes (the dependence in $\ell$ drops out at such large radii)
\be
- \frac{d^2 \Psi}{du^2} + \left( \left[ (N+1)^2 - \frac{1}{4} + \frac{\mu^2}{\kappa^2} \right] \frac{1}{u^2} -1 \right)\Psi = 0\,,
\label{eq:wave_eq_AdS_far_region}
\ee
where $u \equiv \frac{\omega}{\kappa^2 r}$, whose solution can be given in terms of Hankel functions. Since the solution in the intermediary region could only be obtained analytically in the massless case, we now set $\mu=0$.
Noting that $h(r) \simeq r$ for large $r$, the solution for $\Phi_\omega = \Psi/\left(\sqrt{h}\,r^N\right) \propto u^{N+1/2}\Psi$ is
\be
\Phi_\omega(u) = u^{N+1} \left[ \widehat{C}_1 H^{(1)}_{N+1}(u) + \widehat{C}_2 H^{(2)}_{N+1}(u) \right]\,.
\ee
The coefficient $\widehat{C}_1$ controls the asymptotic ``incoming'' part of the solution, while $\widehat{C}_2$ controls the asymptotic ``outgoing'' part.

Now, in the limit $r\rightarrow\infty$ the wavefunction behaves like
\be
\Phi_\omega(r) \longrightarrow \frac{\widehat{C}_1+\widehat{C}_2}{2^{N+1} \Gamma(N+2)} \left( \frac{\omega}{\kappa^2 r} \right)^{2(N+1)} + i (\widehat{C}_2-\widehat{C}_1) \frac{2^{N+1} \Gamma(N+1)}{\pi} \,.
\label{eq:large_r_behavior}
\ee
Comparing this with the solution we found for the intermediate region~\eqref{eq:region2expansionAdS} we obtain
\be
\widehat{C}_2-\widehat{C}_1 = -i\frac{\pi}{2^{N+1} \Gamma(N+1)} A_I \,, \qquad
\widehat{C}_1+\widehat{C}_2 = i\, 2^N \Gamma(N+1) \left( \frac{\kappa^2 r_h}{\omega} \right)^{2N+1} \frac{h(r_h)}{r_h} A_I \,.
\ee
%

\subsubsection*{Greybody factor}

We can now determine the greybody factor.
First we compute the flux at spatial infinity by employing expression~\eqref{eq:JasyAF},
\be
\widehat{J}_{asy} =  \frac{2\Omega_{2N+1} \omega^{2N+2}}{\pi \kappa^{4N+2}} \left( |\widehat{C}_1|^2 - |\widehat{C}_2|^2 \right) 
\equiv \widehat{J}_{in} - \widehat{J}_{out} \,.
\label{eq:JasyAdS}
\ee
It can be verified that $J_{hor}=\widehat{J}_{asy}$, which is a simple consequence of the total flux being conserved.

Finally, we compute the greybody factor:
\be
\gamma(\omega) = \frac{J_{hor}}{\widehat{J}_{in}} = 1- \frac{|\widehat{C}_2|^2}{|\widehat{C}_1|^2} 
= 1- \left| \frac{1-\eta(\omega)}{1+\eta(\omega)} \right|^2 \,,
\label{eq:greybodyAdS}
\ee
where
\be
\eta(\omega) = - \frac{\widehat{C}_2-\widehat{C}_1}{\widehat{C}_1+\widehat{C}_2} = \frac{\pi\, r_h}{2^{2N+1} \left[\Gamma(N+1)\right]^2 h(r_h)} \left( \frac{\omega}{\kappa^2 r_h} \right)^{2N+1}  =
\frac{\pi^{N+2}  \omega^{2N+1}}{2^{2N} \left[\Gamma(N+1)\right]^3 \kappa^{4N+2} {\cal A}_h} \,.
\label{eq:zee}
\ee
The results~\eqref{eq:greybodyAdS} and~\eqref{eq:zee} --- valid for odd spacetime dimensions $D\geq5$ and in the low-frequency regime~\eqref{eq:lowfreqAF} and~\eqref{eq:lowfreqAdS} --- exactly reproduce the expressions derived in~\cite{Harmark:2007jy} when $a=0$, i.e., in the static limit.
Thus, the computation performed above shows that the inclusion of rotation preserves the simple inverse proportionality law between the greybody factor in the low-frequency regime and the area of the event horizon,
\be
\gamma(\omega) \simeq 
\frac{\pi^{N+2}}{2^{2(N-1)} \left[\Gamma(N+1)\right]^3 {\cal A}_h} \left(\frac{\omega}{\kappa^2}\right)^{2N+1} \,.
\label{eq:greybodyAdS_lowfreq}
\ee
%
%
Since the horizon area decreases with spin (see Appendix~\ref{sec:Appendix} for details), we conclude that adding rotation to the black hole solution tends to {\em increase} the greybody factor in the low-frequency limit.

Greybody factors for massless scalar fields emitted from black 3-branes in AdS$_5$ have been studied in Ref.~\cite{Rocha:2009xy}. This is expected to approximate reasonably well the absorption coefficients for large BHs in five-dimensional AdS and indeed the low-frequency results obtained there precisely agree with~\eqref{eq:greybodyAdS_lowfreq} when setting $N=1$.\footnote{Ref.~\cite{Rocha:2009xy} also considered non-radial scalar modes, but their connection to $m\neq0$ modes studied in Section~\ref{sec:GBF_numerics} is not clear.}

\subsection{Small AdS black holes
\label{sec:smallAdS_analytic}}

As shown in Ref.~\cite{Harmark:2007jy}, an alternative matching computation can also be performed for {\em small} black holes in AdS ($\kappa r_h \ll 1$) in the regime specified by
\be
\ho \ll \frac{T_H}{\kappa},\quad \ho \ll \frac{1}{\kappa r_h}, \quad \kappa r_h \ll 1\,.
\label{eq:small_regime}
\ee
This allows us to identify an intermediate region $r_h \ll r \ll 1/\kappa$ which overlaps with region II, where the wave function behaves like~\eqref{eq:region2expansionAdS}. The main interest of such a calculation is that its regime of validity extends to large frequencies $\ho$ for small BHs, since in that case $T_H/\kappa \simeq (\kappa r_h)^{-1} \gg 1$.

To this end, the scalar wave equation~\eqref{eq:wavetortoise} in region III, far away from the horizon, is expressed in terms of the coordinate $z = \frac{\kappa^2 r^2}{1 + \kappa^2 r^2}$ as (here we reinstate factors of $\ell$)
\be
\left[4 z (1- z) \frac{d^2 }{dz^2} + 2 (1-2z)\frac{d}{dz} + \ho^2- \frac{(N+1/2)(N-1/2+2z)}{z (1-z)} - \frac{\ell(\ell+2N)}{z}\right](r^{\frac{2N+1}{2}}\Phi_\omega) = 0\,.
\label{eq:HyperEDO}
\ee
For generic choices of the frequency, the general solution to this equation is given by
\beq
r^{\frac{2N+1}{2}}\Phi_\omega &=& C_1 z^{\frac{2N+1+2\ell}{4}}(1-z)^{-\frac{2N+1}{4}} {}_2F_1\Big(\frac{\ho+\ell}{2},-\frac{\ho-\ell}{2},N+1+\ell,z\Big) \nonumber\\
&& + \, C_2 z^{-\frac{2N-1+2\ell}{4}}(1-z)^{\frac{2N+3}{4}} {}_2F_1\Big(1-\frac{\ho+\ell}{2},1+\frac{\ho-\ell}{2},N+2,1-z \Big)\,.
\label{eq:solsHyper}
\eeq
However, note that the hypergeometric differential equation~\eqref{eq:HyperEDO} degenerates whenever
\be
\ho=2(N+1)+\ell+2p\,, \qquad p=0,1,2,\dots
\label{eq:AdSnormalfrequencies}
\ee
in which case the two solutions given in~\eqref{eq:solsHyper} cease to be linearly independent. Therefore, this computation breaks down at this discrete set of frequencies.

Now we must take the limit $z \to 0$ to compare with region II. For $\ell=0$ this yields the same result as in Ref.~\cite{Harmark:2007jy} and comparing it with~\eqref{eq:region2expansionAdS} we obtain
\be
C_1 = A_I \kappa^{-\frac{2N+1}{2}}, \qquad C_2 = A_I \frac{i \omega \, r_h^{2N}h(r_h)}{2N}\frac{\kappa^{\frac{2N-1}{2}}\Gamma\left(N+1+\frac{\ho}{2}\right)\Gamma\left(N+1-\frac{\ho}{2}\right)}{\Gamma(N)\Gamma\left(N+2\right)}\,.
\label{eq:matchCs}
\ee

Taking instead the limit $z \to 1$ with $1-z\approx\frac{1}{\kappa^2 r^2}$ (using formula 15.3.11 from Ref.~\cite{Abramowitz:1970as}), we get the approximate wave function
\beq
\Phi_\omega &=& C_1 \frac{\kappa^{\frac{2N+1}{2}}\Gamma(N+1+\ell)\Gamma(N+1)}{\Gamma\left(N+1+\frac{\ho+\ell}{2}\right)\Gamma\left(N+1-\frac{\ho-\ell}{2}\right)} + \dots \\ 
&& + \frac{1}{\kappa^{\frac{2N+3}{2}}r^{2N+2}}\Big[C_2 + C_1 {\frac{\Gamma(N+1+\ell)}{ \Gamma(N+2) \Gamma\left(\frac{\ho+\ell}{2}\right)\Gamma\left(-\frac{\ho-\ell}{2}\right)}} \alpha(\ell) \Big] + \dots\,, \nonumber
\eeq
where 
\be
\alpha(\ell) = \psi(1) + \psi(N+2) - \psi\left(N+1-\frac{\ho-\ell}{2}\right) - \psi\left(N+1+\frac{\ho+\ell}{2}\right)\,,
\ee
and we recall $\psi(x)$ represents the digamma function. Note the term proportional to $C_1$ in the second line: this term contributes at the same order (in $r$) as the term proportional to $C_2$ and must be included for consistency, a detail that was overlooked in Ref.~\cite{Harmark:2007jy}. Correcting this we no longer find frequencies of total reflection or transmission.

The leading terms from the expansion of the Hankel functions in the general solution for $\kappa r \gg 1$ were given in Eq.~\eqref{eq:large_r_behavior}. Matching with this region fixes the coefficients $\widehat{C}_1$ and $\widehat{C}_2$ in terms of the coefficients $C_1$ and $C_2$,
\beq
\widehat{C}_2-\widehat{C}_1 &=&  C_1 \frac{\pi \kappa^{\frac{2N+1}{2}} \Gamma(N+1+\ell)}{i2^{N+1}\Gamma\left(N+1+\frac{\ho+\ell}{2}\right)\Gamma\left(N+1-\frac{\ho-\ell}{2}\right)}\,, \\
\widehat{C}_1 + \widehat{C}_2 &=& \frac{2^{N+1}\kappa^{\frac{6N+5}{2}}\Gamma(N+2)}{\omega^{2N+2}}\Big\{ C_2 \\
&& \!\!\!\!\!\!\!\!\!\!\!\!\!\!\!\!\!\!\!\!\!\!\!\!   + C_1 \frac{\Gamma(N+1+\ell)}{\Gamma(N+2) \Gamma\left(\frac{\ho+\ell}{2}\right)\Gamma\left(-\frac{\ho-\ell}{2}\right)}\Big[\psi(1)+\psi(N+2)-\psi\left(N+1-\frac{\ho-\ell}{2}\right)-\psi\left(N+1+\frac{\ho+\ell}{2}\right)\Big] \Big\}. \nonumber
\eeq
Restricting to $\ell=0$ modes and using relations~\eqref{eq:matchCs} we obtain
\be
\widehat{C}_2-\widehat{C}_1 =  A_I \frac{\pi \Gamma(N+1)}{i2^{N+1}\Gamma\left(N+1+\frac{\ho}{2}\right)\Gamma\left(N+1-\frac{\ho}{2}\right)}\,,
\ee
and
\beq
\widehat{C}_1 + \widehat{C}_2 &=& A_I \frac{\kappa^{\frac{6N+5}{2}}\Gamma(N+2)2^{N+1}}{\omega^{2N+2}}\Big[ \frac{\psi(1)+\psi(N+2)-\psi(N+1-\ho/2)-\psi(N+1+\ho/2)}{\kappa^{\frac{2N+1}{2}} (N+1) \Gamma\left(\frac{\ho}{2}\right)\Gamma\left(-\frac{\ho}{2}\right)} \nonumber\\
&& + \frac{i \omega h(r_h) r_h^{2N}}{2N}\frac{\kappa^{\frac{2N-1}{2}}\Gamma\left(N+1+\frac{\ho}{2}\right)\Gamma\left(N+1-\frac{\ho}{2}\right)}{\Gamma(N)\Gamma(N+2)}\Big].
\eeq
Taking the quotient of these two expressions, we get the quantity $\eta(\omega) =\frac{\widehat{C}_1-\widehat{C}_2}{\widehat{C}_1 + \widehat{C}_2}$, related to the greybody factor through
\be
\gamma(\omega)=1-\left|\frac{1-\eta(\ho)}{1+\eta(\ho)}\right|^2,
\label{eq:AdS_lf}
\ee
where
\beq
\eta(\ho) &=& \frac{\pi\, \ho^{2N+2}}{2^{2N+2} (N+1)\Gamma\left(N+1+\frac{\ho}{2}\right)\Gamma\left(N+1-\frac{\ho}{2}\right)} \nonumber\\
&& \times \frac{1}{\left[i \frac{\psi(N+1+\ho/2) + \psi(N+1-\ho/2) - \psi(1) - \psi(N+2)}{(N+1)\Gamma\left(\frac{\ho}{2}\right)\Gamma\left(-\frac{\ho}{2}\right)} + \ho\frac{ h(r_h)r_h^{2N}\kappa^{2N+1}\Gamma\left(N+1+\frac{\ho}{2}\right)\Gamma\left(N+1-\frac{\ho}{2}\right)}{2\,\Gamma(N+1)\Gamma(N+2)}\right]}\,.
\label{eq:eta_small}
\eeq
In the low-frequency regime, $\ho \ll 1$, this expression reduces to Eq.~\eqref{eq:zee}, which constitutes a good check on our calculation.

The formulas obtained above are not very enlightening but a simpler expression can be derived for frequencies large compared with the dimensionless horizon radius,
\be
\kappa r_h \ll \ho \ll \frac{1}{\kappa r_h}\,.
\label{eq:freq_reg}
\ee
The second condition comes from~\eqref{eq:small_regime}. Under this assumption, the real part in the denominator of~\eqref{eq:eta_small} becomes negligible. Defining $Z\equiv Re(\eta^{-1})$ and $W\equiv Im(\eta^{-1})$, we may then expand around $Z=0$, thus obtaining
\beq
\gamma(\omega) &=& 1-\frac{|\eta^{-1}-1|^2}{|\eta^{-1}+1|^2} = 1-\frac{(Z-1)^2+W^2}{(Z+1)^2+W^2} \nonumber\\
&& \!\!\!\!\!\!\!\!\!\!\!\!\!\!\!\!\!\!\! \simeq \frac{4Z}{1+W^2} = \frac{2^{2N+2}\kappa^{2N+1} {\cal A}_h\, \Gamma(N+1+\ho/2)^2 \Gamma(N+1-\ho/2)^2}{\pi^{N+2} \ho^{2N+1} \Gamma(N+1)^3 \left\{ 1+\left(\frac{2^{2N+2}\Gamma(N+1+\ho/2)\Gamma(N+1-\ho/2)}{\pi \ho^{2N+2} \left[\psi(1)+\psi(N+2)-\psi(N+1+\ho/2)-\psi(N+1-\ho/2) \right] }\right)^2\right\}}\,.
\label{eq:amazingformula}
\eeq
Note that the quantity $W$ depends solely on the dimensionality $N$, besides the frequency $\ho$. The only dependence on the black hole itself, namely on the mass and spin, enters through $Z$ and, in particular, the horizon area. Furthermore,  this dependence on the black hole horizon area is linear, at least in the frequency regime~\eqref{eq:freq_reg}. The greybody factor features large-amplitude oscillations, while interpolating between $0$ at low frequencies and $1$ at high frequencies. Nevertheless, this intricate structure is universal, in the sense that, once the dimensionality of the spacetime is fixed, the greybody factor is the same for any asymptotically AdS small black hole, up to overall scaling.

\begin{figure}[t]
\centering
\subfigure{\includegraphics[width=8.9cm]{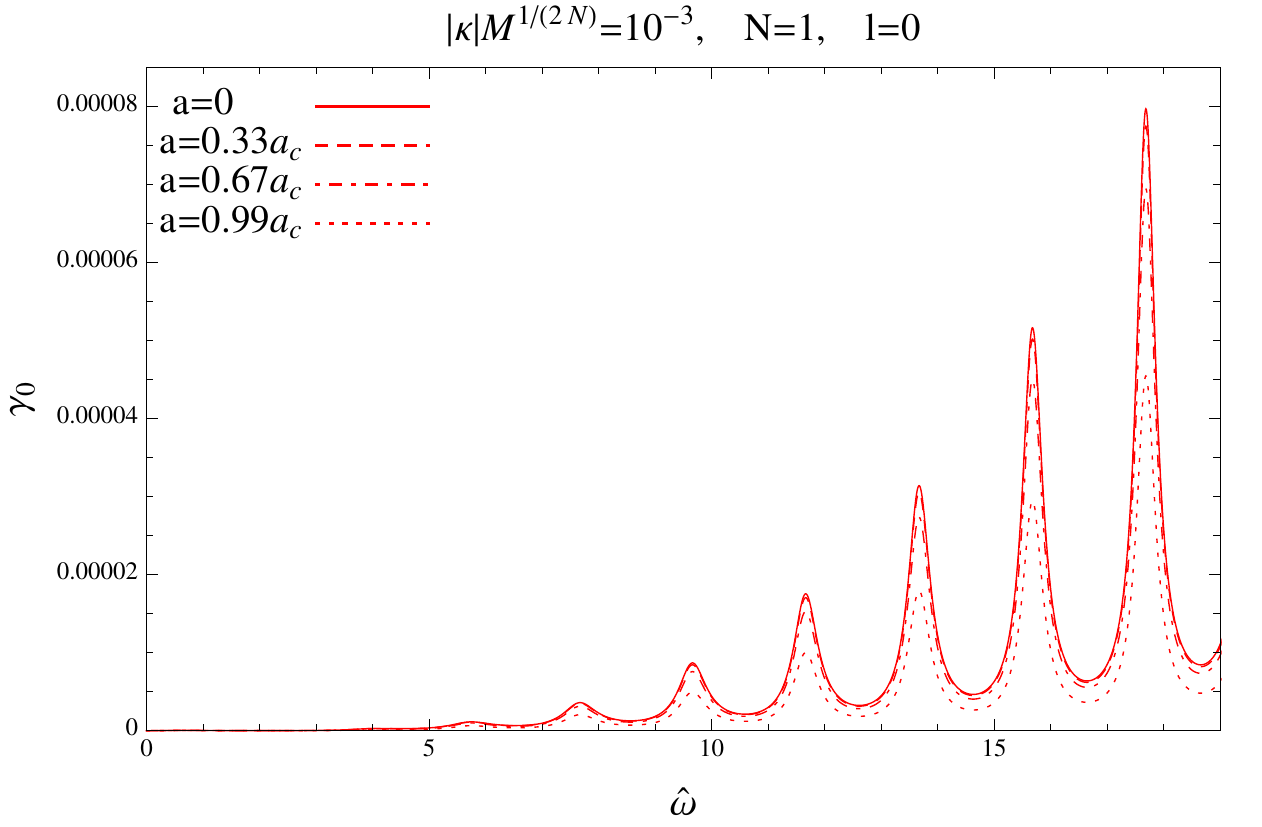}}
\subfigure{\includegraphics[width=8.9cm]{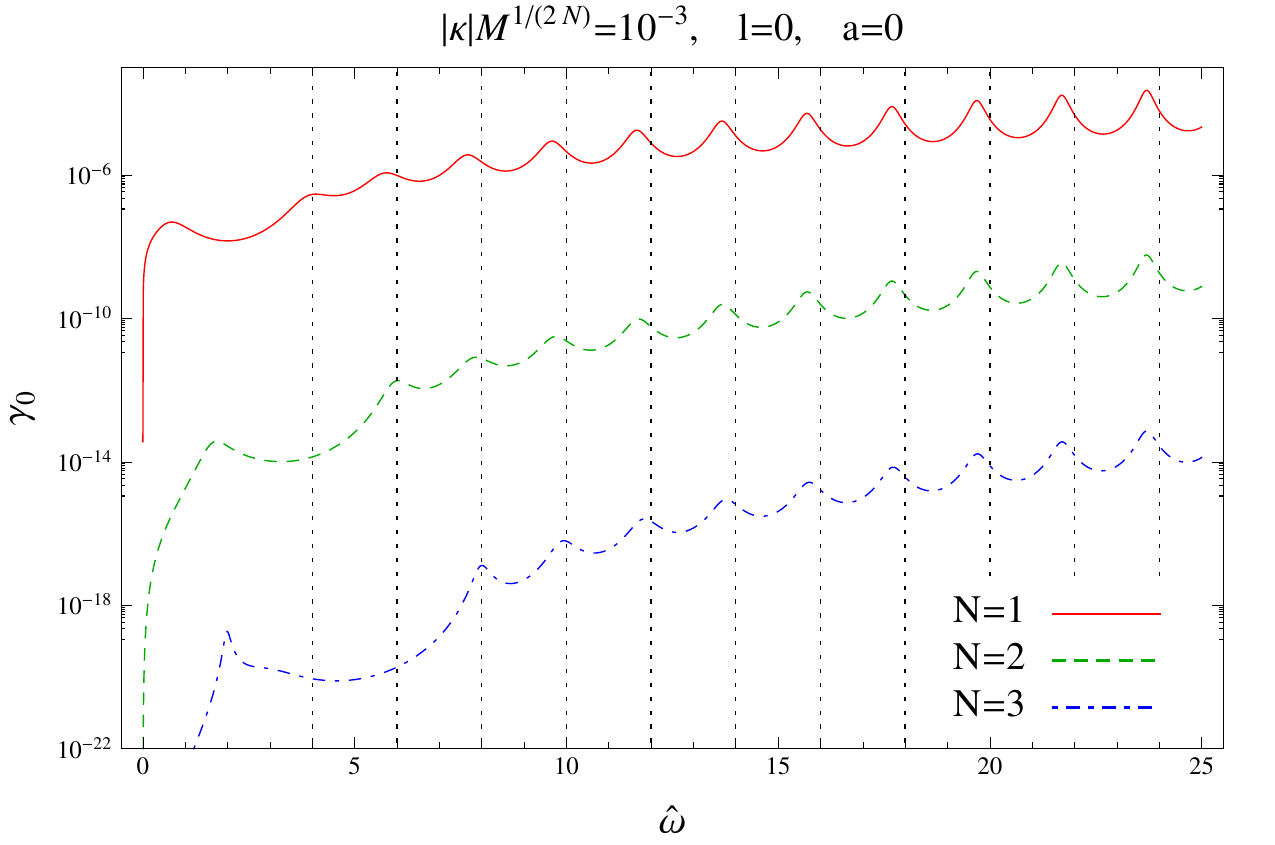}}
\caption{Greybody factors for small asymptotically AdS black holes and $\ell=0$ modes, as determined by formula~\eqref{eq:amazingformula}, which is accurate for frequencies $\kappa r_h \ll \ho \ll (\kappa r_h)^{-1}$. {\it Left panel:} $N=1$, mass parameter $\kappa M^{1/(2N)}=10^{-3}$, and several rotation parameters $a/a_c=0, 0.33, 0.67, 0.99$. {\it Right panel:} Same as left panel but fixing $a=0$ and varying the dimensionality, $N=1,2,3$. The vertical dotted lines mark the location of the normal modes of pure AdS$_5$. (The normal modes for the cases $N=2$ and $N=3$ are exactly the same, except that the lowest normal frequency occurs at $\ho=6$ for $N=2$, and at $\ho=8$ for $N=3$.)}
\label{fig:gbf_small_AdS_ana}
\end{figure}

Given the complexity of the expressions above, our results for small AdS black holes can be interpreted more easily by visual inspection. In Fig.~\ref{fig:gbf_small_AdS_ana} we show the greybody factor determined by the analytical formula~\eqref{eq:amazingformula}.\footnote{Note that expression~\eqref{eq:amazingformula} is regular at the normal frequencies of pure AdS$_{2N+3}$, which are given by Eq.~\eqref{eq:AdSnormalfrequencies}. This is in contrast with formula~\eqref{eq:eta_small}, which breaks down precisely at those points, as we discussed. Apart from this, the two results~\eqref{eq:AdSnormalfrequencies} and~\eqref{eq:amazingformula} are in excellent agreement over the frequency range defined by~\eqref{eq:freq_reg}, for the class of small AdS black holes.} 
The greybody spectrum shows large-amplitude oscillations, with a periodicity roughly consistent with the spacing between the normal modes of pure AdS spacetime. While Ref.~\cite{Harmark:2007jy} pointed out that such normal modes coincided with critical frequencies of full reflection (i.e. vanishing greybody factor), our improved analysis indicates that they correspond more closely to the peaks of the spectrum. At any rate, the relation between maximum or minimum transmission frequencies and normal modes of pure AdS is unclear, especially because the boundary conditions that are imposed at the black hole horizon (which, for small black holes, is close to $r=0$) are different from the ones demanded at the origin for pure AdS.

\mysection{Numerical evaluations and comparison with analytic results
\label{sec:GBF_numerics}}

In this section we determine numerically the greybody factor for a massless scalar field in rotating cohomo\-geneity-1 black hole spacetimes and compare with our previous analytic results. The numerical evaluation includes the modes with $m\neq0$, which are not amenable to analytic treatment along the lines of Section~\ref{sec:GBF_analytic}. For $m>0$ and low frequencies such that $\omega < m \Omega_h$ the phenomenon of superradiance is expected to occur~\cite{Zeldovich:1972,Press:1972zz}, meaning that the reflected wave has larger amplitude than the incident wave. This translates into the corresponding greybody factor becoming negative for small $\omega$.

It can be shown that the scattering coefficients for $m=0$ modes are the same if we consider, instead of the absorption of an impinging wave, the process of Hawking emission from the black hole horizon~\cite{Harmark:2007jy}. On the other hand, if $m\neq0$ the scattering coefficients for the two processes differ, as we now briefly review.

First, we shall need to generalize some equations obtained in Section~\ref{sec:GBF_analytic} for modes with $m\neq0$. Let us begin by analyzing the asymptotically flat case. As before, the potential~\eqref{eq:potential} vanishes at the location of the event horizon but for $m\neq0$ the wave equation~\eqref{eq:wavetortoise} receives a correction. This correction is absent as $r\to\infty$ since then $\Omega(r)\to0$. The wavefunction thus has the following behaviors
\be
\Psi^{+} \sim \left\{ 
\begin{array}{l}
B^{+}\, e^{-i \wto x} \qquad\qquad\quad \text{as} \;\; x \to -\infty\,, \\
e^{-i\omega x} + A^{+}\, e^{i\omega x} \qquad\, \text{as} \;\; x \to \infty\,,
\end{array}
\right.
\label{eq:scat_absorption}
\ee
where $\wto \equiv \omega-m \Omega_h$. The solution above satisfies the boundary conditions appropriate for the process in which a scalar wave impinges on the black hole, part of it being absorbed and part being reflected. In the opposite situation of spontaneous emission of scalar waves by the black hole horizon the wavefunction behaves as
\be
\Psi^{-} \sim \left\{ 
\begin{array}{l}
e^{i \wto x} + A^{-}\, e^{-i \wto x} \qquad \text{as} \;\; x \to -\infty\,, \\
B^{-}\, e^{i\omega x} \qquad\qquad\quad\;\; \text{as} \;\; x \to \infty\,.
\end{array}
\right.
\label{eq:scat_emission}
\ee
The quantities $|A^{\pm}|^2$ and $|B^{\pm}|^2$ represent reflection and transmission coefficients, respectively, and are functions of $\omega, \ell$ and $m$ in general.

By using the constancy of the Wronskian between pairs of solutions of the wave equation and considering various combinations of the solutions~\eqref{eq:scat_absorption}, \eqref{eq:scat_emission} and their complex conjugates it is possible to derive important relations~\cite{DeWitt:1975ys,Ottewill:2000qh}:
\beq
\omega B^{-} &=& \wto B^{+}\,, \label{eq:relations1}\\
-\wto A^{-} {B^{+}}^* &=& \omega {A^{+}}^* B^{-}\,, \\
1-|A^{-}|^2 &=& \frac{\omega}{\wto} |B^{-}|^2\,, \\
1-|A^{+}|^2 &=& \frac{\wto}{\omega} |B^{+}|^2\,. \label{eq:relations4}
\eeq
These relations are not all independent and also imply
\be
|A^{-}|^2 = |A^{+}|^2\,.
\label{eq:equal_reflection_coefs}
\ee

As in Section~\ref{sec:GBF_analytic}, the greybody factor for a scalar wave impinging on a black hole from infinity, $\gamma_{abs}$ is computed by the ratio between the flux at the horizon and the incoming part of the asymptotic flux,
\be
\gamma_{abs} = \frac{J_{hor}}{J_{in}} = \frac{\wto}{\omega} |B^{+}|^2 = 1 - |A^{+}|^2\,.
\label{eq:GBF_abs}
\ee
The last equality makes manifest that the greybody factor is bounded from above by unity. The next-to-last equality also shows that the greybody factor becomes negative whenever $\wto<0$, i.e. in the superradiant regime.

Considering instead the process of Hawking emission from the black hole horizon, for which the appropriate boundary conditions are those in Eq.~\eqref{eq:scat_emission}, the greybody factor is determined by the ratio between the asymptotic flux and the outgoing part of the flux at the horizon,
\be
\gamma_{emit} = \frac{J_{asy}}{J_{out}} = \frac{\omega}{\wto} |B^{-}|^2 = 1 - |A^{-}|^2\,.
\ee
The identity~\eqref{eq:equal_reflection_coefs} then shows that the greybody factors for these two different physical processes are equal,
\be
\gamma_{abs} = \gamma_{emit} \equiv \gamma\,,
\ee
just like in the non-rotating limit~\cite{Harmark:2007jy}. The same cannot be said of the transmission coefficients $|B^{\pm}|^2$, which coincide only when $\wto=\omega$, or equivalently when $m\Omega_h=0$. Finally, we point out that the transmission coefficients $|B^{\pm}|^2$ need not be bounded from above by one and generically there will be a frequency range over which they exceed unity.

Everything until now concerned the asymptotically flat case. For the asymptotically dS black holes the effective potential $V(r)$ also vanishes at both ends of the integration domain (in this case at $r=r_h$ and $r=r_c$) so the formal solutions near the horizons are same as above but there is one difference: the cosmological horizon also has a finite angular velocity and therefore the solutions describing the absorption of impinging scalar waves and spontaneous emission are respectively
\be
\Psi^{+} \sim \left\{ 
\begin{array}{l}
B^{+}\, e^{-i \wto x} \qquad\qquad\quad \text{as} \;\; x \to -\infty\,, \\
e^{-i\omega' x} + A^{+}\, e^{i\omega' x} \quad\;\,\, \text{as} \;\; x \to \infty\,,
\end{array}
\right.
\qquad\quad
\Psi^{-} \sim \left\{ 
\begin{array}{l}
e^{i \wto x} + A^{-}\, e^{-i \wto x} \qquad \text{as} \;\; x \to -\infty\,, \\
B^{-}\, e^{i\omega' x} \qquad\qquad\quad\; \text{as} \;\; x \to \infty\,,
\end{array}
\right.
\ee
where $\omega' \equiv \omega-m \Omega_c$. Thus we obtain the same relations as in the asymptotically flat case~(\ref{eq:relations1}--\ref{eq:relations4}), with the only difference that we should make the replacement $\omega\to\omega'$.

Finally we turn to the asymptotically AdS case. Here the situation is significantly different from the asymptotically flat and dS geometries because the effective potential does not vanish at spatial infinity. In fact, there it diverges as $r^2$, corresponding to the boundary $x=0$ in terms of the tortoise coordinate. The wave equation near $x=0$ is independent of $m$ and given by Eq.~\eqref{eq:wave_eq_AdS_far_region}, whose general solution is
\be
\Psi \sim \sqrt{-\omega x} \left[ \widehat{C}_1 H^{(1)}_{N+1}(-\omega x) + \widehat{C}_2 H^{(2)}_{N+1}(-\omega x) \right] \qquad\qquad \text{for}\;\; x\sim0 \,.
\ee
Recall that the coefficient $\widehat{C}_1$ controls the asymptotic ``incoming'' part of the solution, while $\widehat{C}_2$ determines the ``outgoing'' flux. It is now useful to expand in a power series around $x=0$, as was done in~\eqref{eq:large_r_behavior}. After doing so and conveniently rescaling the coefficients $\widehat{C}_1$ and $\widehat{C}_2$ we arrive at
\beq
\Psi^{+} &\sim& \left\{ 
\begin{array}{l}
B^{+}\, e^{-i \wto x} \qquad\qquad\qquad\qquad\qquad\qquad\qquad\qquad\qquad\quad\;\; \text{as} \;\; x \to -\infty\,, \\
\sqrt{\frac{\pi}{2}} \left[ (1+A^{+})\frac{(-\omega x)^{N+3/2}}{2^{N+1}\Gamma(N+2)} + i(A^{+}-1)\frac{2^{N+1}\Gamma(N+1)}{\pi(-\omega x)^{N+1/2}} \right]  \quad\; \text{as} \;\; x \to 0\,,
\end{array}
\right. 
\\
\, \nonumber \\
\Psi^{-} &\sim& \left\{ 
\begin{array}{l}
e^{i \wto x} + A^{-}\, e^{-i \wto x} \qquad\qquad\qquad\qquad\qquad\quad\;\,\; \text{as} \;\; x \to -\infty\,, \\
\sqrt{\frac{\pi}{2}} \left[ B^{-}\frac{(-\omega x)^{N+3/2}}{2^{N+1}\Gamma(N+2)} + i B^{-}\frac{2^{N+1}\Gamma(N+1)}{\pi(-\omega x)^{N+1/2}} \right]  \quad\; \text{as} \;\; x \to 0\,.
\end{array}
\right.
\eeq
Interestingly, precisely the same relations as in the flat case, Eqs.~(\ref{eq:relations1}--\ref{eq:relations4}), are obtained by considering once again the constancy of the Wronskian acting on pairs of solutions with the above asymptotic behaviors.

Thus, we have just shown that the relations~(\ref{eq:relations1}--\ref{eq:relations4}) between the various scattering coefficients hold generically for odd dimensional, cohomogeneity-1, Myers-Perry geometries, irrespective of the value of the cosmological constant, with the understanding that the frequency $\omega$ should be replaced by $\omega - m \Omega(r_o)$, where $r_o$ is the outer boundary of the region covered by the tortoise coordinate.

To perform the numerical evaluation of the greybody factors for the asymptotically flat and dS black holes, in practice, we start by imposing the boundary condition $\Psi = e^{-i\wto x}$ at very large negative values of $x$ corresponding to the black hole event horizon ($x\to-\infty$). Then we integrate numerically Eq.~\eqref{eq:scalarwave} up to very large positive values of $x$ (or up to $x=0$ in the asymptotically AdS case) where we match the solution to
\be
\Psi^{+} \sim \left\{
\begin{array}{l}
  \frac{1}{B^{+}}e^{-i\omega x} + \frac{A^{+}}{B^{+}}e^{i\omega x} \hspace{4.9cm} \text{as}\;\; x\to\infty \;\; \text{in asymptotically flat case}\,, \\ \, \\
  \frac{1}{B^{+}}e^{-i(\omega-m\Omega_c) x} + \frac{A^{+}}{B^{+}}e^{i(\omega-m\Omega_c) x} \hspace{2.8cm} \text{as}\;\; x\to\infty \;\; \text{in asymptotically dS case}\,.
\end{array}
\right.
\ee
From this we extract the coefficients $A^{+}$ and $B^{+}$ and compute the greybody factor using Eq.~\eqref{eq:GBF_abs}. We monitor the accuracy of our results by verifying that
\be
|A^{+}|^2 + \frac{\omega-m\Omega(r_h)}{\omega-m\Omega(r_o)} \, |B^{+}|^2=1\,.
\ee
In the case of asymptotically AdS black holes, results are obtained by matching the numerical solution of Eq.~\eqref{eq:scalarwave} with the asymptotic solution~\eqref{eq:solsHyper}. Numerical errors are estimated by the violation in the flux conservation, i.e., by comparing Eq.~\eqref{eq:Jhor} with Eq.~\eqref{eq:JasyAdS}. 
The flux conservation violations of the results presented never exceed $0.01\,\%$ and in addition convergence tests were also performed. In all cases, the numerical solution to Eq.~\eqref{eq:scalarwave} is obtained via a fourth-fifth order Runge-Kutta method.

We present below a selection of numerical results, including comparisons with our analytic findings. The parameters presented in the figures were conveniently chosen so to exhibit the key features of the greybody factors. However, the  applied methods may be generalized to obtain results for any parameter configuration of a scalar field propagating around a rotating black hole in a cohomogeneity-1 spacetime with cosmological constant.

\subsection{Asymptotically flat black holes}

\begin{figure}[t]
\centering
\includegraphics[width=17cm]{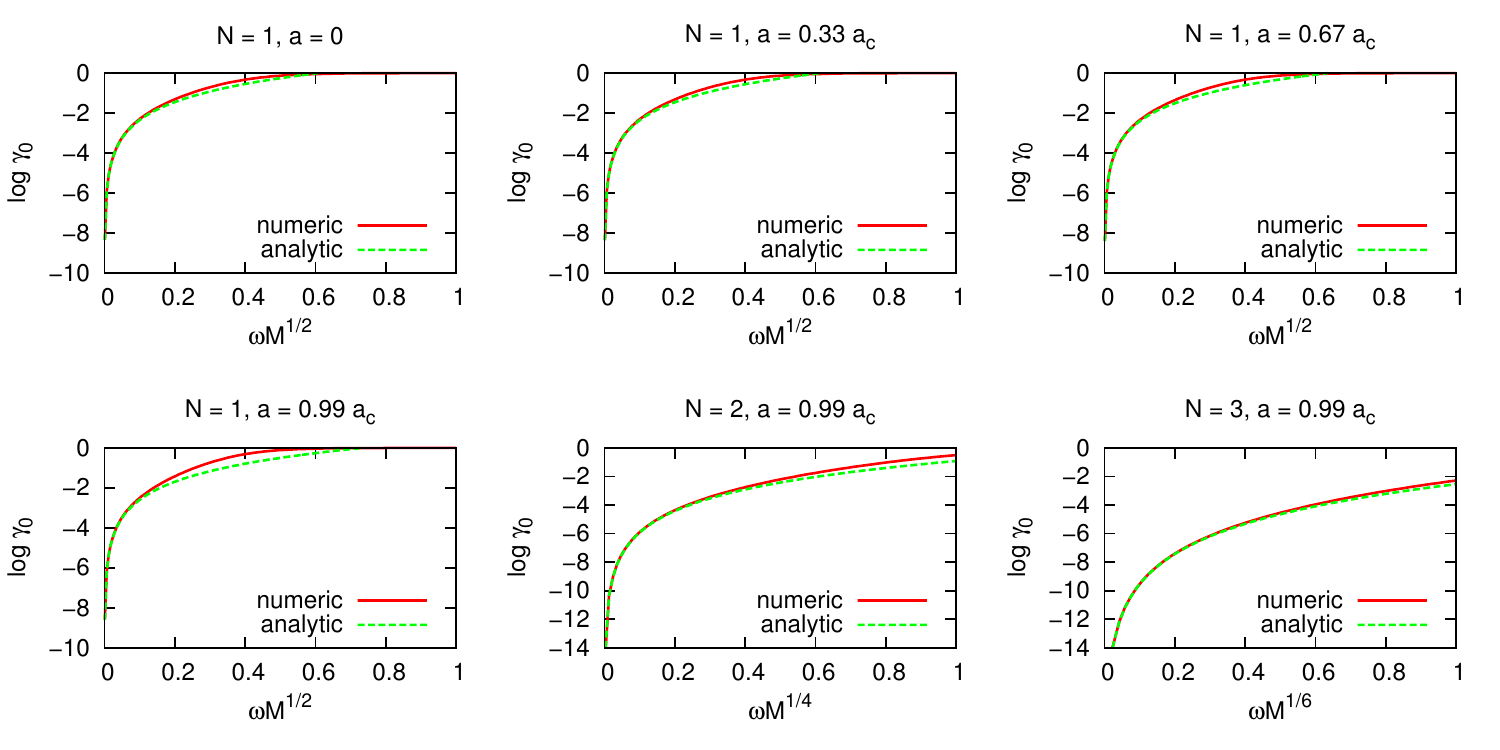}
\caption{Comparison between analytic and numerical results for greybody factors in the asymptotically flat case, restricting to $\ell=m=0$ modes. A selection of dimensions ($N=1,2,3$) and rotations ($a/a_c = 0, 0.33, 0.67, 0.99$) are shown.}
\label{fig:af_comp}
\end{figure}

\begin{figure}[H]
\centering
\includegraphics[width=12cm]{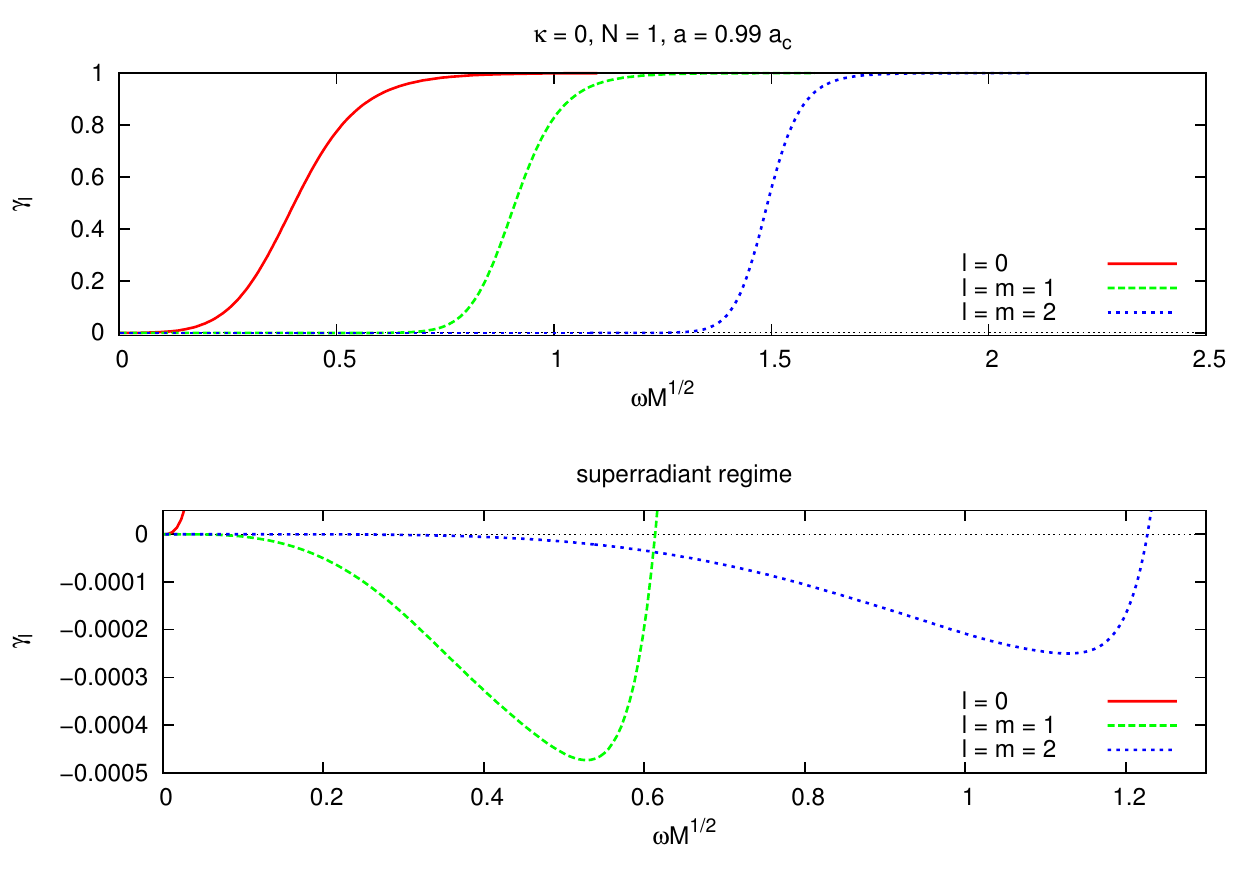}
\caption{{\it Top panel:} Numerical results for greybody factors in the asymptotically flat case, for modes with $\ell=m=0,1,2$. For this figure the values $N=1$ and $a=0.99\,a_c$ were adopted. {\it Bottom panel:} Enlarged reproduction of the top panel, focusing on the superradiant regime, where the greybody factors become negative.}
\label{fig:af_N1a0.99}
\end{figure}

\begin{figure}[H]
\centering
\includegraphics[width=12cm]{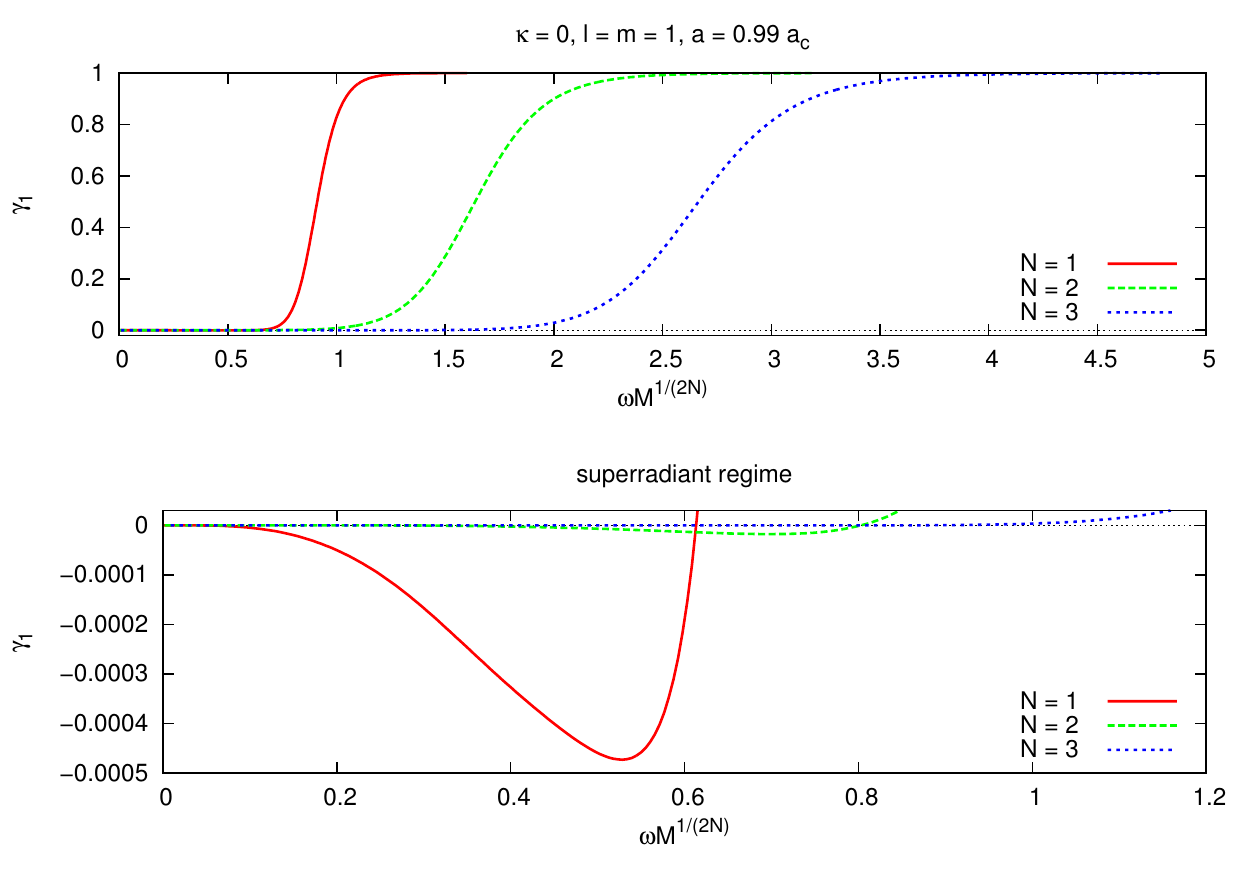}
\caption{{\it Top panel:} Numerical results for greybody factors in the asymptotically flat case, for various dimensions $N=1,2,3$. In this figure only modes $\ell=m=1$ and $a=0.99\,a_c$ were considered. {\it Bottom panel:} Enlarged reproduction of the top panel, focusing on the superradiant regime.}
\label{fig:AF_l1}
\end{figure}

\begin{figure}[t]
\centering
\includegraphics[width=11cm]{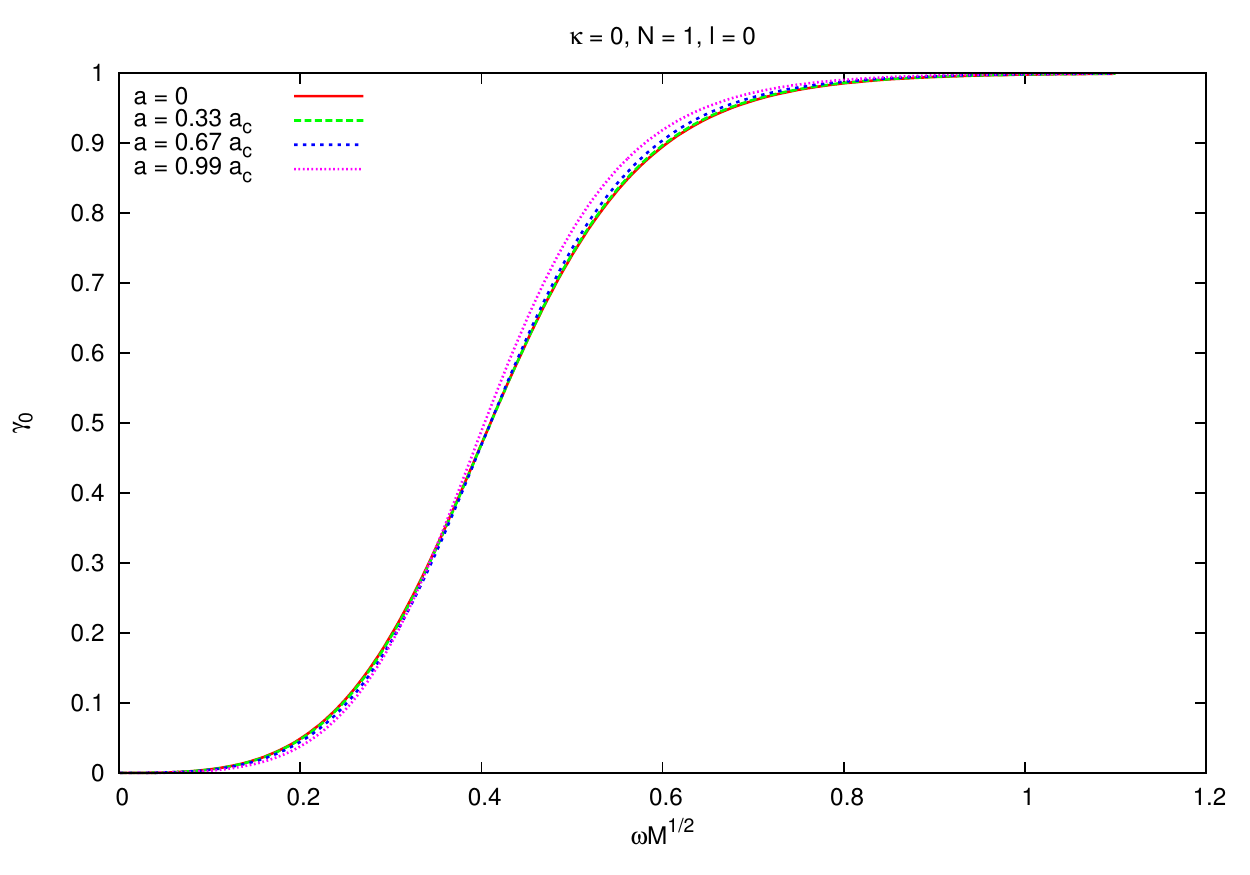}
\caption{Numerical results for greybody factors in the asymptotically flat case, for $N=1$, $\ell=0$ and various spin parameters, $a/a_c = 0, 0.33, 0.67, 0.99$. The dependence on the rotation is weak, and similar behavior was observed in the cases $N=2,3$.}
\label{fig:af_N1l0}
\end{figure}

Figures~\ref{fig:af_comp}--\ref{fig:af_N1l0} show our numerical results for asymptotically flat black holes ($\kappa = 0$). In Fig.~\ref{fig:af_comp}, numerical results are compared with their analytic equivalents valid in the low-frequency regime, Eq.~\eqref{eq:af-low_freq_gbf}.  The comparisons were made for $\ell = m = 0$, $N = 1$ with $a = 0, 0.33\, a_c, 0.67 \, a_c, 0.99 \, a_c$, and $a = 0.99 \, a_c$ with $N = 2, 3$. We note that in all cases the analytic results agree very well with the numerical ones in their regime of validity, which is $\omega r_h \ll 1$.

We show greybody factors for $N = 1$, $a = 0.99 \, a_c$, and $ \ell = m = 0, 1, 2$ in Fig.~\ref{fig:af_N1a0.99}. While in the top panel we present an overall comparison, in the bottom panel we zoom in on the superradiant regime. As expected, there is no superradiance for the $ \ell = 0$ mode and it is most efficient for near maximal spin parameter, $a=0.99a_c$, and $\ell = m = 1$. The superradiant regime is also shown in Fig.~\ref{fig:AF_l1}, but now for a fixed value of $\ell$ and $m$ in three different dimensions. As can be seen, superradiance is highly suppressed as the number of dimensions increases. In the superradiant regime, the amplification factor is commonly defined as the excess of the reflection coefficient compared to the incidence coefficient, expressed in percentage:
\be
\text{amplification factor} = 100\,(|A^+|^2-1)\,\%\,.
\ee
The results we present in all our figures are not for the reflection coefficient $|A^+|^2$ but instead for the greybody factor. Nevertheless, one can use expression~\eqref{eq:GBF_abs} to convert between these two quantities. Upon doing so, we obtain maximum amplification factors --- corresponding to minimum greybody factors --- equal to $0.046\,\%$ for $N = 1$, $0.0017\,\%$ for $N = 2$, and $3.1\times 10^{-5}\,\%$ for $N = 3$. Note that, in the asymptotically flat case, these values are independent of the mass parameter $M$.\footnote{The amplification factor obviously depends on the choice of rotation parameter, $a$. To obtain the numerical values presented we considered $a=0.99\, a_c$\,.}

Figure~\ref{fig:af_N1l0} shows how the greybody factor for the $\ell = 0$ mode is affected by the change in rotation of five-dimensional black holes ($N = 1$). We conclude that the spin has a weak influence on the greybody factor for $\ell = 0$.

\subsection{de Sitter black holes}

\begin{figure}[H]
\centering
\includegraphics[width=16.5cm]{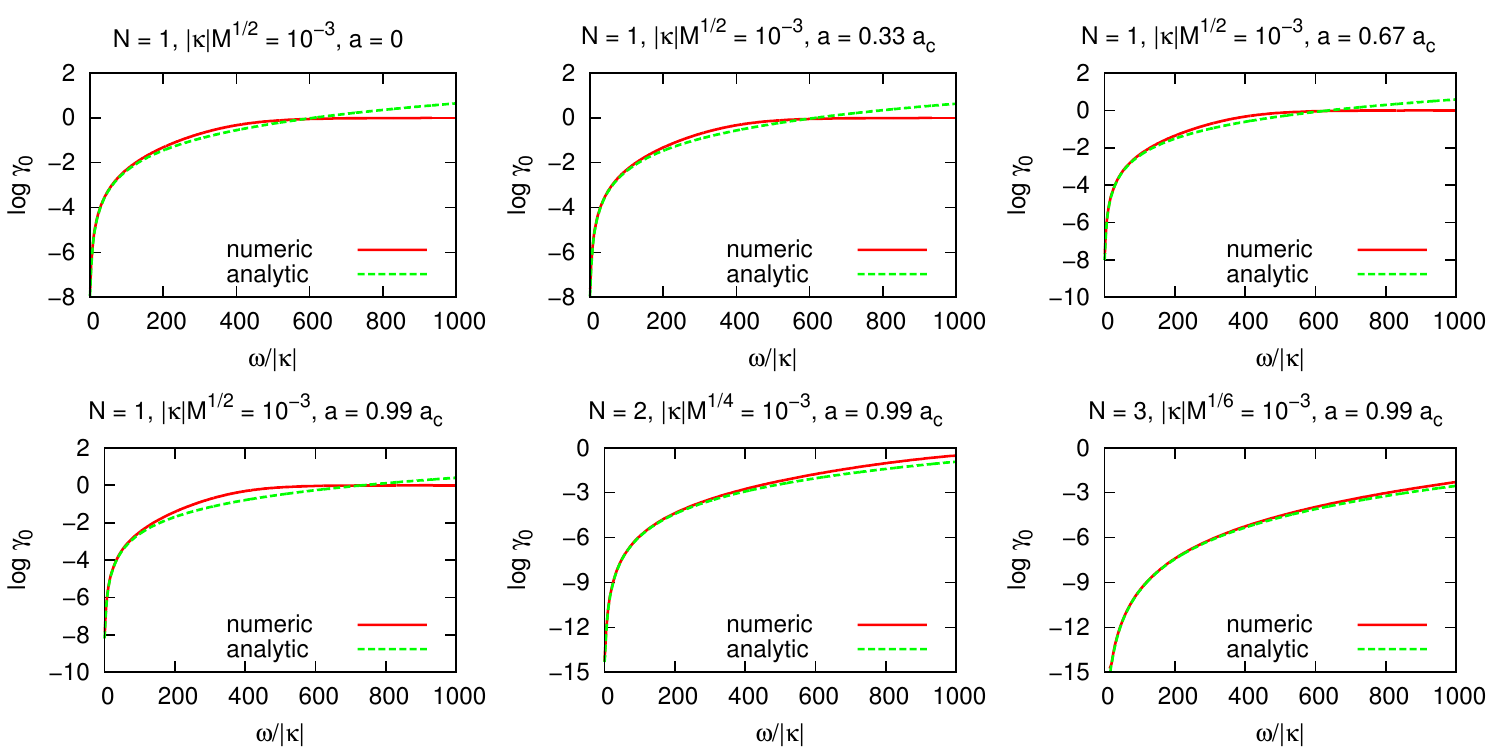}
\caption{Comparison between analytic and numerical results for greybody factors in the asymptotically de Sitter case, restricting to $\ell=m=0$ modes. A selection of dimensions ($N=1,2,3$) and rotations ($a/a_c = 0, 0.33, 0.67, 0.99$) are shown and the choice of mass parameter $|\kappa|M^{1/2N}=10^{-3}$ was made, corresponding to the class of small de Sitter black holes.}
\label{fig:dS_comp}
\end{figure}

\begin{figure}[H]
\centering
\includegraphics[width=10.5cm]{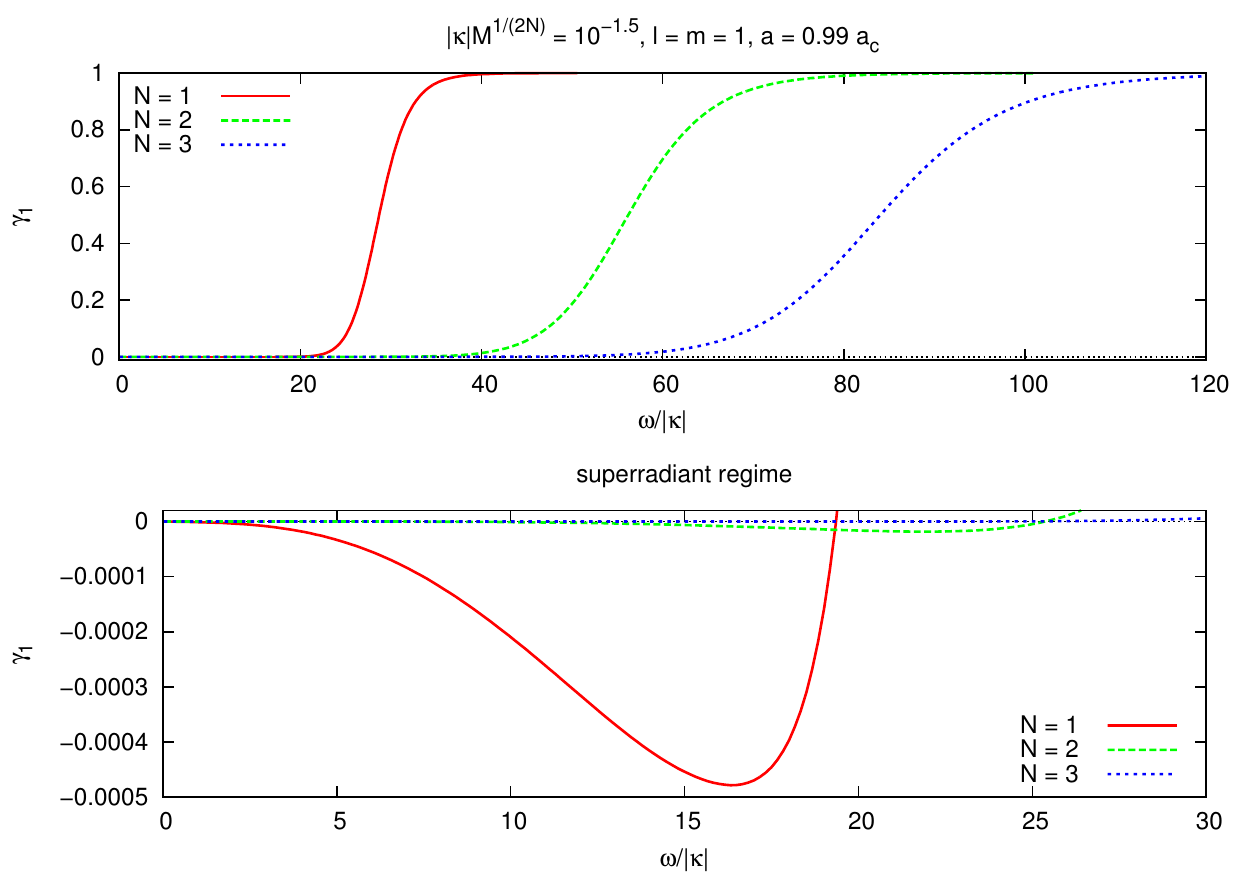}
\caption{{\it Top panel:} Greybody factors (numerical) for asymptotically de Sitter small black holes, for various dimensions $N=1,2,3$. In this figure only modes $\ell=m=1$ and $a=0.99\,a_c$ were considered and the choice of mass parameter $|\kappa|M^{1/2N}=10^{-1.5}$ was made. {\it Bottom panel:} Enlarged reproduction of the top panel, focusing on the superradiant regime. These results are quantitatively similar to those of asymptotically flat black holes (Fig.~\ref{fig:AF_l1}), with the understanding that the horizontal axis must be rescaled by a factor of $|\kappa|M^{1/2N}=10^{-1.5}$.}
\label{fig:dS_small-l1a0.99}
\end{figure}

\begin{figure}[H]
\centering
\subfigure{\includegraphics[width=8.9cm]{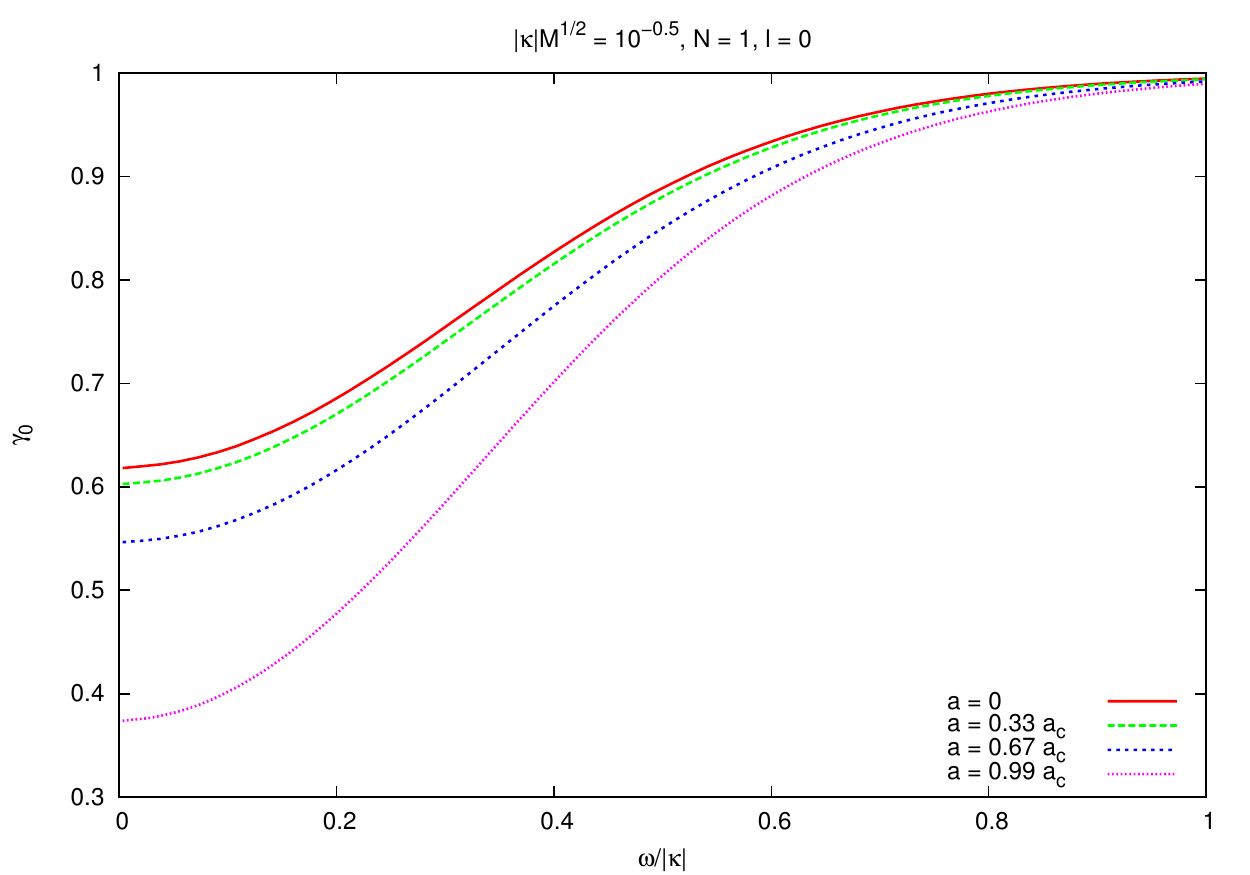}}
\subfigure{\includegraphics[width=8.9cm]{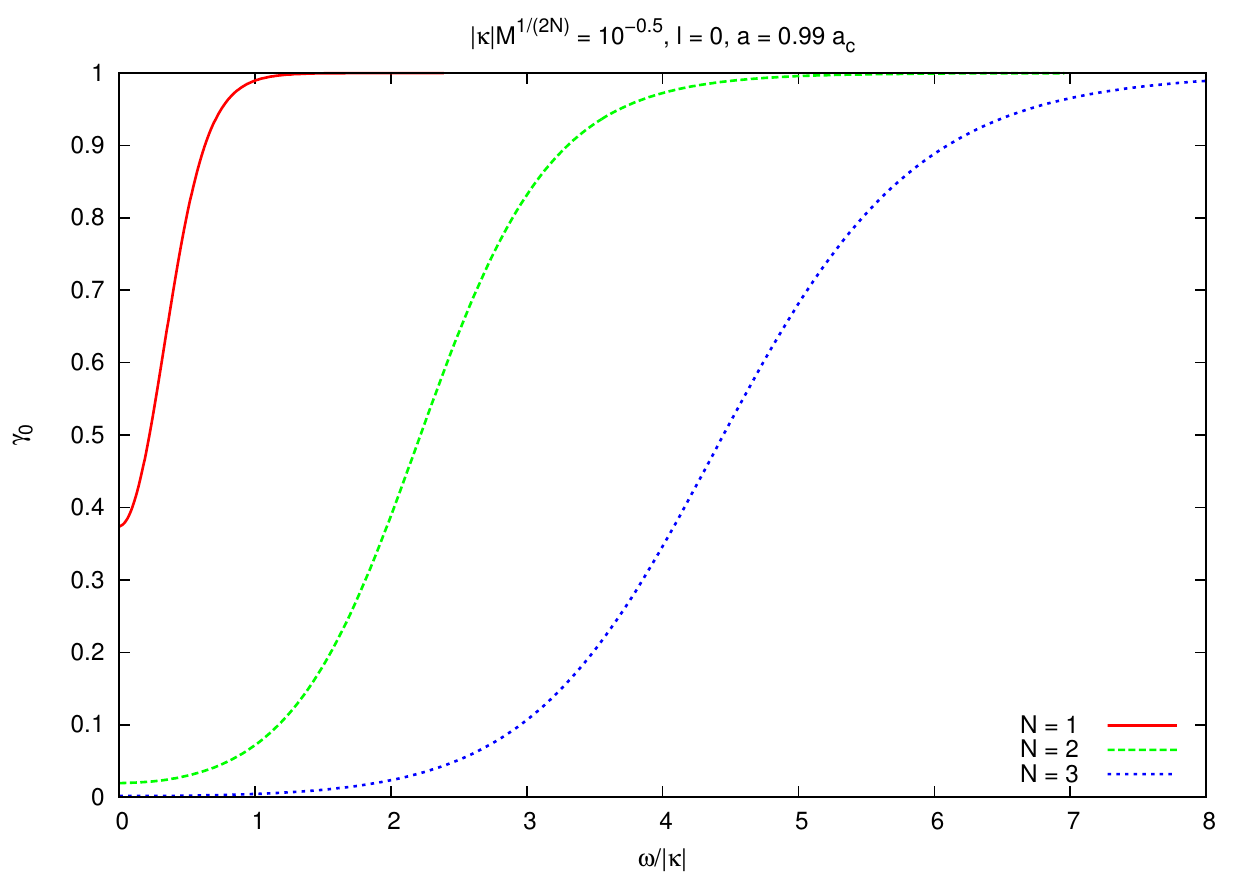}}
\caption{{\it Left panel:} Greybody factors (numerical) for asymptotically de Sitter large black holes, for $N=1$, $\ell=m=0$, mass parameter $|\kappa|M^{1/2N}=10^{-0.5}$ and various rotation parameters, $a/a_c=0, 0.33, 067, 0.99$. {\it Right panel:} Same as left panel but fixing $a=0.99 a_c$ and varying the dimensionality, $N=1,2,3$.}
\label{fig:dS_big-l0}
\end{figure}

\begin{figure}[H]
\centering
\includegraphics[width=11cm]{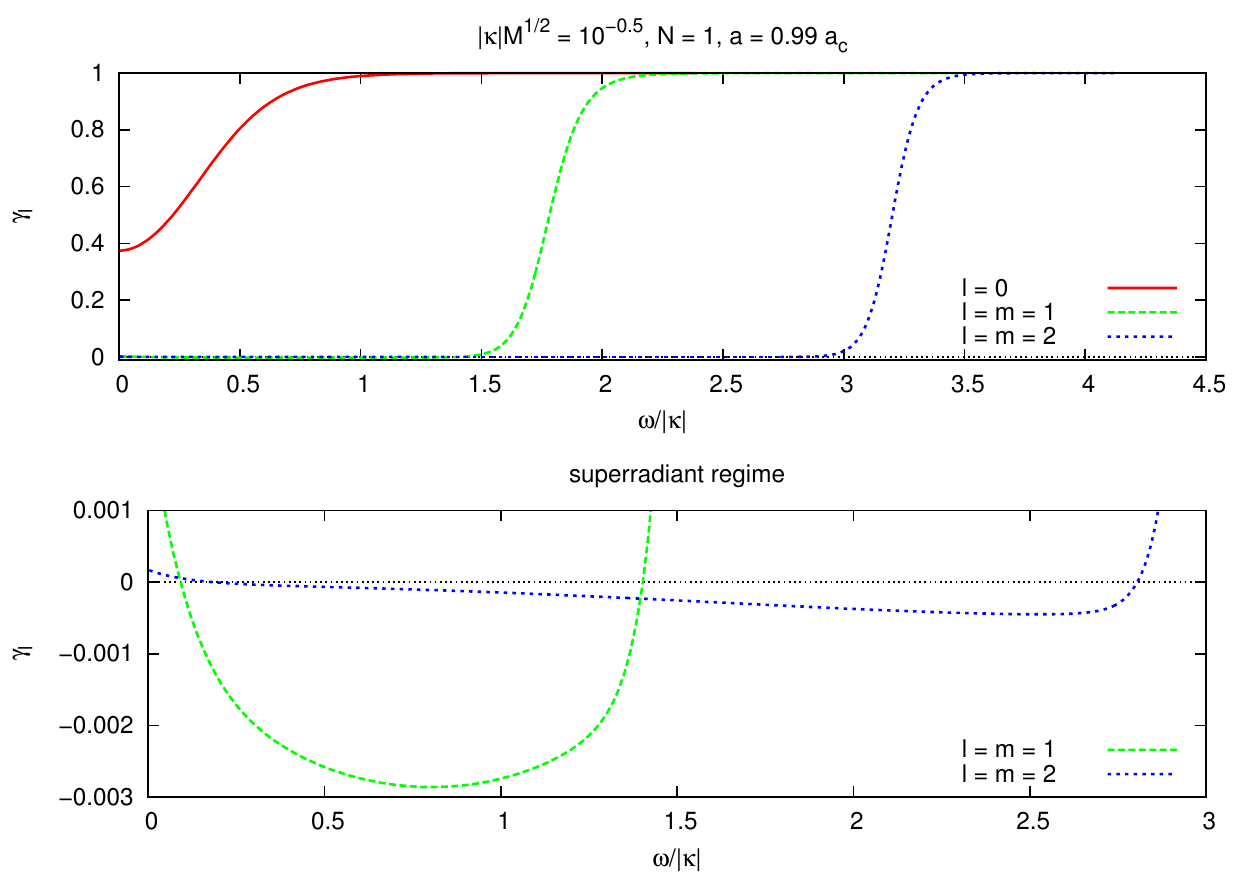}
\caption{{\it Top panel:} Greybody factors (numerical) for asymptotically de Sitter large black holes, for modes with $\ell=m=0,1,2$. For this figure the values $|\kappa|M^{1/2}=10^{-0.5}$, $N=1$ and $a=0.99\,a_c$ were adopted. {\it Bottom panel:} Enlarged reproduction of the top panel, focusing on the superradiant regime.}
\label{fig:dS_big-N1a0.99}
\end{figure}

\begin{figure}[t]
\centering
\includegraphics[width=11cm]{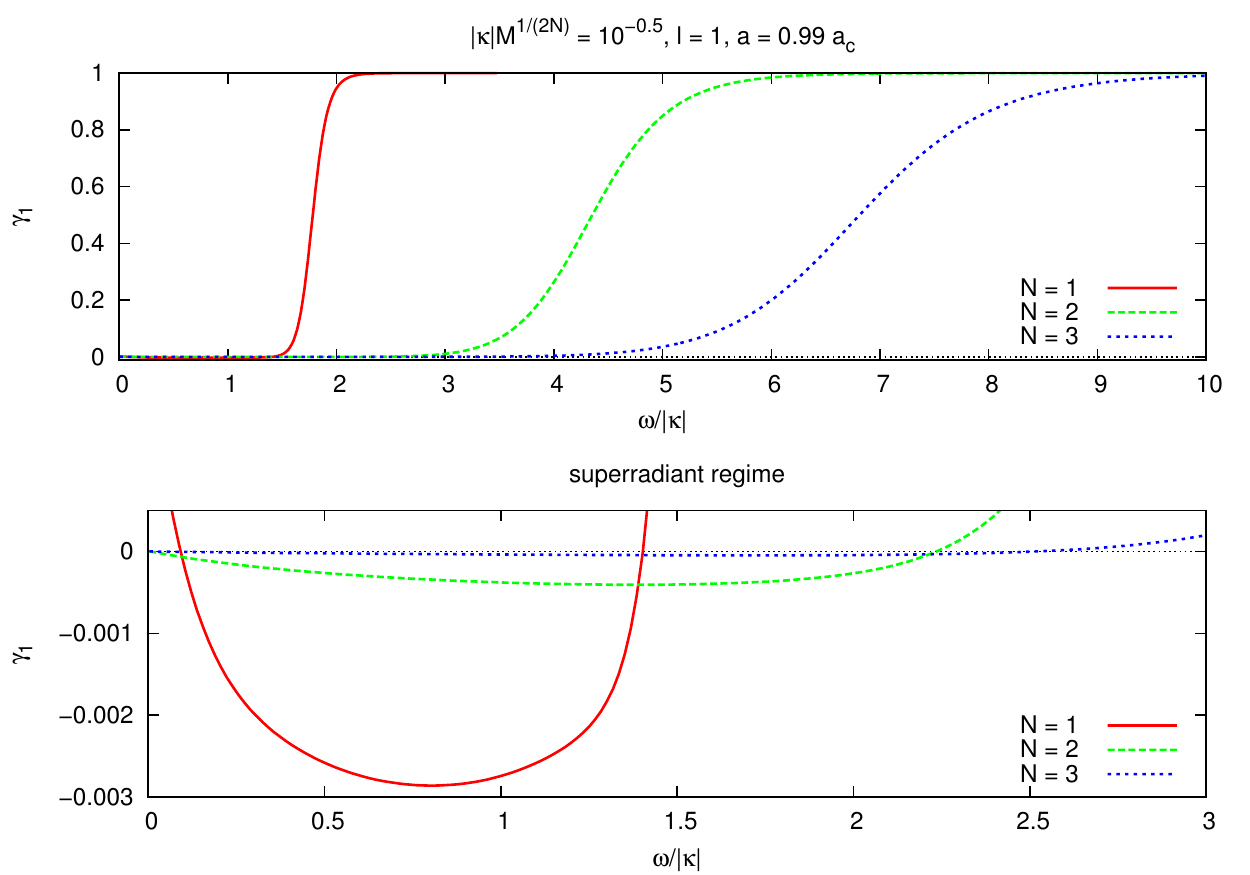}
\caption{{\it Top panel:} Greybody factors (numerical) for asymptotically de Sitter large black holes, for various dimensions $N=1,2,3$ and mass parameter $|\kappa|M^{1/2N}=10^{-0.5}$. In this figure only modes $\ell=m=1$ and $a=0.99\,a_c$ were considered. {\it Bottom panel:} Enlarged reproduction of the top panel, focusing on the superradiant regime.}
\label{fig:dS_big-l1a0.99}
\end{figure}

\begin{figure}[H]
\centering
\includegraphics[width=11cm]{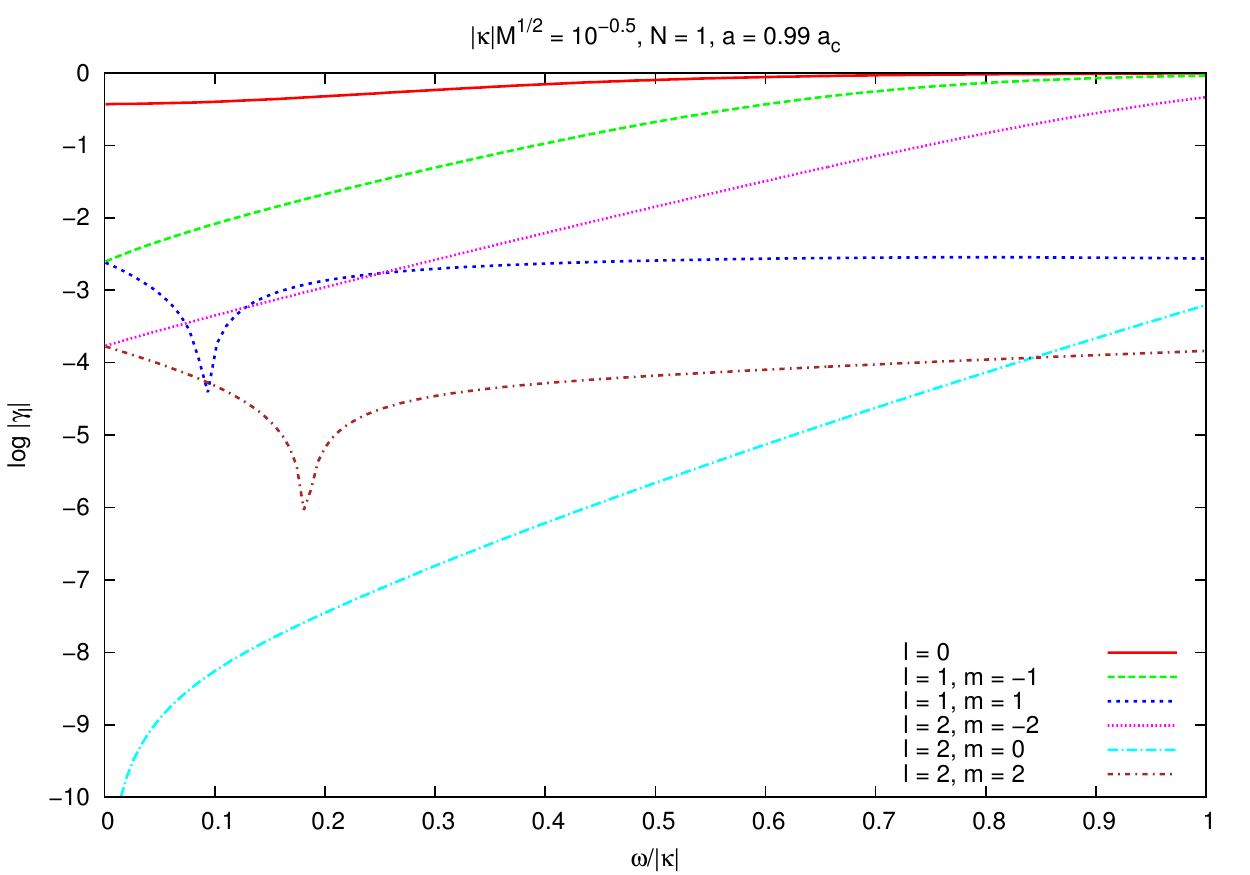}
\caption{Greybody factors (numerical) for asymptotically de Sitter large black holes, for $N=1$, mass parameter $|\kappa|M^{1/2}=10^{-0.5}$, rotation parameter $a=0.99\,a_c$ and $\ell=0,1,2$ for the possible choices of $m$. A logarithmic vertical scale is employed. Note that for $m>0$ modes the greybody factor becomes negative at small frequencies and therefore it is the absolute value that is plotted.}
\label{fig:dS_big-N1a0.99log}
\end{figure}

Results for dS black holes are presented in Figs.~\ref{fig:dS_comp}--\ref{fig:dS_big-N1a0.99log}. In Fig.~\ref{fig:dS_comp} we compare numerical results with the analytic result for small black holes ($r_h \ll r_c$), Eq.~\eqref{eq:GBF_dS}, which is valid for $\ell = 0$ and low frequencies ($\omega \ll \min \{ \,|\kappa|, T_H\}$). We have set $|\kappa| M^{1/(2N)} = 10^{-3}$, which falls within the regime of small dS black holes, and we present results for $N = 1, 2, 3$ and a selection of sub-extremal rotation parameters, $a/a_c<1$. In the low-frequency regime, the agreement between numerical and analytic results is excellent, independently of the chosen configuration.

Numerical results for $\ell = m = 1$, $a = 0.99 \, a_c$, $N = 1, 2, 3$, and $|\kappa|M^{1/(2N)} = 10^{-1.5}$ are shown in Fig.~\ref{fig:dS_small-l1a0.99}, corresponding to the case of a small black hole in dS. The top panel displays the overall comparison, while the bottom panel shows a zoom-in on the superradiant regime. As in the asymptotically flat case, superradiance is suppressed when the number of dimensions increases. We get $0.048\,\%$ of maximum amplification factor in the $N = 1$ case, $0.0018\,\%$ for $N = 2$, and $3.4\times 10^{-5}\,\%$ for $N = 3$.

In Fig.~\ref{fig:dS_big-l0} we compare different results for the mode $\ell = m = 0$ and $|\kappa|M^{1/(2N)} = 10^{-0.5}$, corresponding to the case of a large black hole in dS. The left graph shows greybody factors for $N = 1$ and different BH rotation values. The right graph shows results for $a = 0.99 \, a_c$ in different dimensions ($N = 1, 2, 3$). It is worth noting that the greybody factors tend to a nonzero value in the $\omega \to 0$ limit. This is also valid for small black holes but more noticeable for large values of $|\kappa| M^{1/(2N)}$. A non-vanishing zero-frequency greybody factor has been already demonstrated to occur for Schwarzschild--de Sitter black holes in four-~\cite{Brady:1996za} and higher-dimensional~\cite{Kanti:2005ja} spacetimes. This conclusion, however, only holds if the scalar field is minimally coupled to the spacetime curvature. For generic couplings, the greybody factors tend to zero in the zero-frequency limit~\cite{Crispino:2013pya, Kanti:2014dxa}.

A comparison between modes with $\ell = m = 0, 1, 2$ for $N = 1$, $a = 0.99 \, a_c$, and $|\kappa| M^{1/2} = 10^{-0.5}$ is presented in Fig.~\ref{fig:dS_big-N1a0.99}. While the top panel shows the overall comparison, the bottom panel focuses on the superradiant regime. Like in all cases presented so far, the mode with largest superradiant rate is the $\ell = m = 1$ mode. The maximum amplification factor is $0.29\,\%$ for $\ell = m = 1$ and $0.045\,\%$ for $\ell = m = 2$. Comparing this with the result for small black holes, $|\kappa|M^{1/2} = 10^{-1.5}$ (\emph{cf.} Fig.~\ref{fig:dS_small-l1a0.99}), we conclude that superradiance in dS is enhanced when larger black holes are considered.

The strongest superradiant mode, $\ell = m = 1$, is analyzed in Fig.~\ref{fig:dS_big-l1a0.99} for $a = 0.99 \, a_c$ and $|\kappa|M^{1/(2N)} = 10^{-0.5}$ in different dimensions. As it happens for small black holes (\emph{cf.} Fig.~\ref{fig:dS_small-l1a0.99}), superradiance is suppressed as the number of dimensions is increased. Whereas for $N=1$ the maximum superradiance is $0.29\,\%$, for $N=2$ its value decreases to $0.041\,\%$, and to $0.0048\,\%$ for $N = 3$. 

From the analysis of Fig.~\ref{fig:dS_big-l1a0.99}, there are two points which are noteworthy: (i) the superradiant regime is now restricted to frequencies $m\Omega(r_c) < \omega < m\Omega(r_h)$ instead $0 < \omega < m\Omega(r_h)$, as it was in the asymptotically flat case; (ii) the greybody factor is nonzero in the zero-frequency limit for $ m \neq 0$ modes. Case (i) originates from the fact that spacetime rotation affects both the black hole event and cosmological horizons, although it is stronger in the former. In order to help visualize point (ii), Fig.~\ref{fig:dS_big-N1a0.99log} presents the log plots of the results obtained for $N = 1$, $|\kappa| M^{1/2} = 10^{-0.5}$, $a = 0.99 \, a_c$, $\ell = 1$ ($m = -1,1$), and $\ell = 2$ ($m = -2, 0, 2$). For $m = \pm 1, \pm 2$, the greybody factor is nonzero in the limit $\omega \to 0$, while it is zero for $m = 0$ in the same limit. Therefore, there is nonzero flux in the event horizon of zero-frequency modes that have angular momentum components along the black hole rotation axis. This phenomenon is active only for rotating black holes, since the zero-frequency limit gives a vanishing greybody factor for these modes when $a\to0$, in accordance with~\cite{Kanti:2014dxa}.  As mentioned before, greybody factors are nonzero in the zero-frequency limit for $\ell = 0$ modes regarding positive cosmological constant and minimally-coupled scalar fields~\cite{Brady:1996za,Kanti:2005ja,Crispino:2013pya, Kanti:2014dxa}.

\subsection{Anti-de Sitter black holes}

\begin{figure}[t]
\centering
\includegraphics[width=17cm]{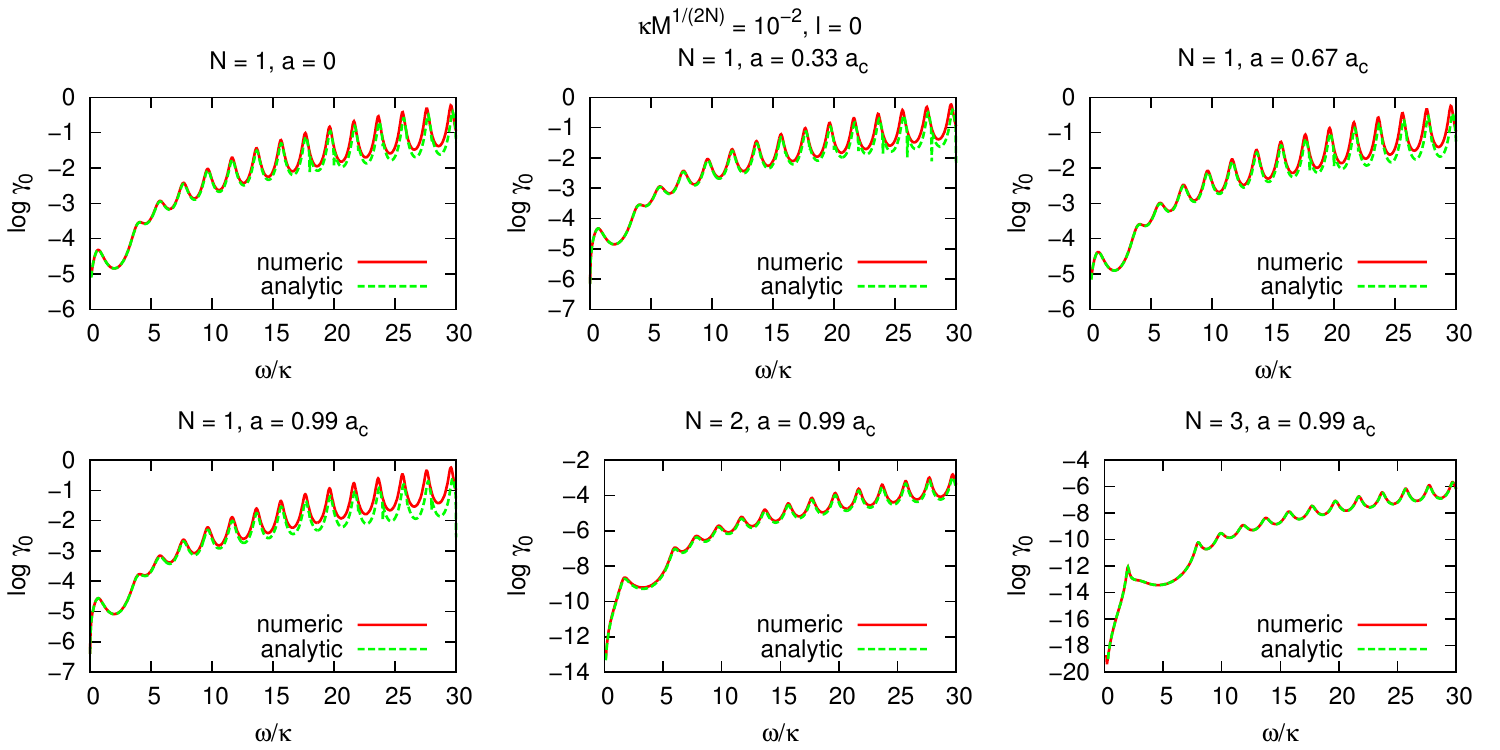}
\caption{Comparison between analytic and numerical results for greybody factors in the asymptotically AdS case, restricting to $\ell=m=0$ modes. A selection of dimensions ($N=1,2,3$) and rotations ($a/a_c = 0, 0.33, 0.67, 0.99$) are shown and the choice of mass parameter $|\kappa|M^{1/2N}=10^{-2}$ was made, corresponding to the class of small AdS black holes.}
\label{fig:AdS_comp}
\end{figure}

\begin{figure}[H]
\centering
\subfigure{\includegraphics[width=8.9cm]{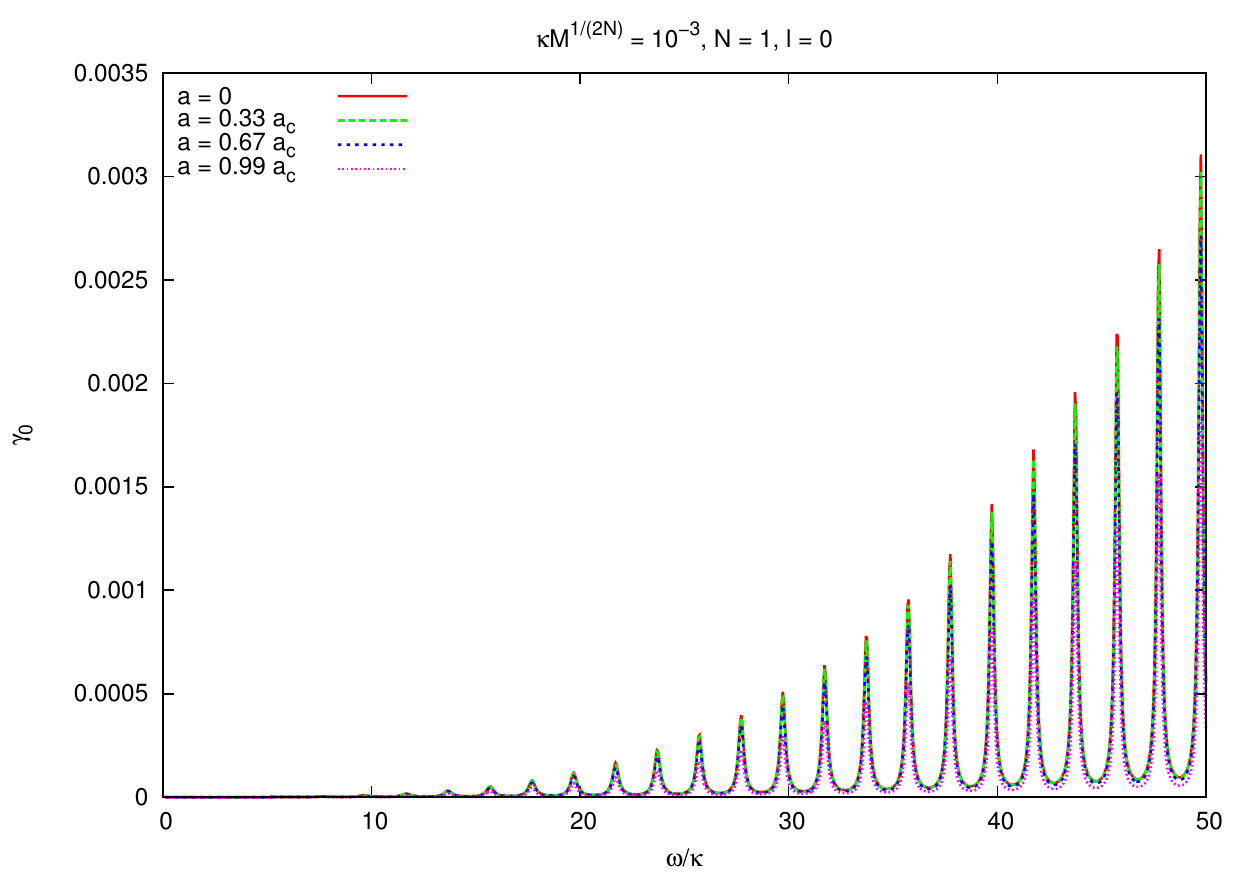}}
\subfigure{\includegraphics[width=8.9cm]{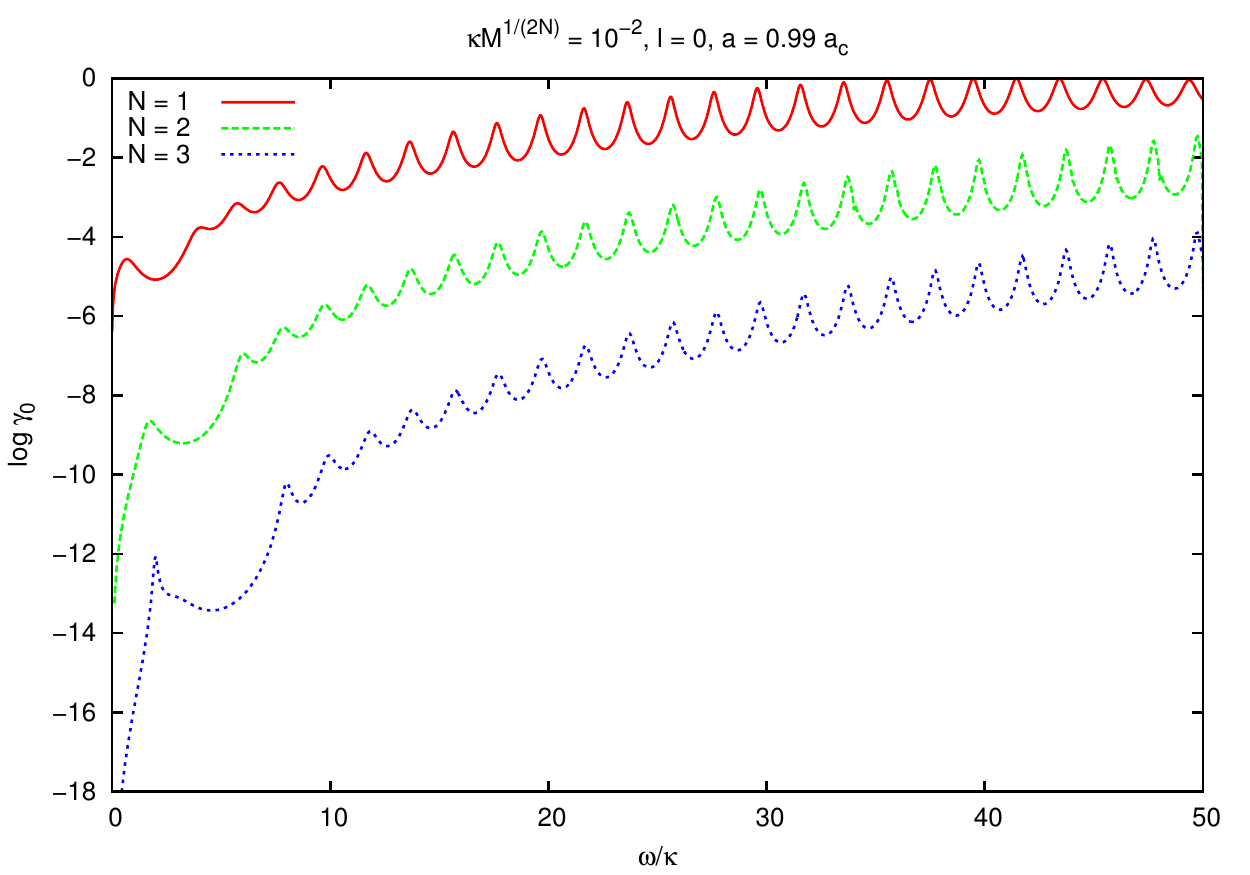}}
\caption{Greybody factors (numerical) for small asymptotically Anti--de Sitter black holes, for $ \ell = 0$ modes. The structure of these results is identical to those presented in Fig.~\ref{fig:gbf_small_AdS_ana}. {\it Left panel:} A selection of results for $\kappa M^{1/2} = 10^{-3}$, $N=1$ and various choices of the spin parameter are presented ($a/a_c=0,\, 0.33,\, 0.67,\, 0.99$). {\it Right panel:} Same as left panel, except now we have set $\kappa M^{1/(2N)} = 10^{-2}$ and $a = 0.99 a_c$, varying the specetime dimensionality ($N = 1,2,3$) and using a vertical log scale.}
\label{fig:AdS_small-l0a0.99}
\end{figure}

\begin{figure}[H]
\centering
\includegraphics[width=11.5cm]{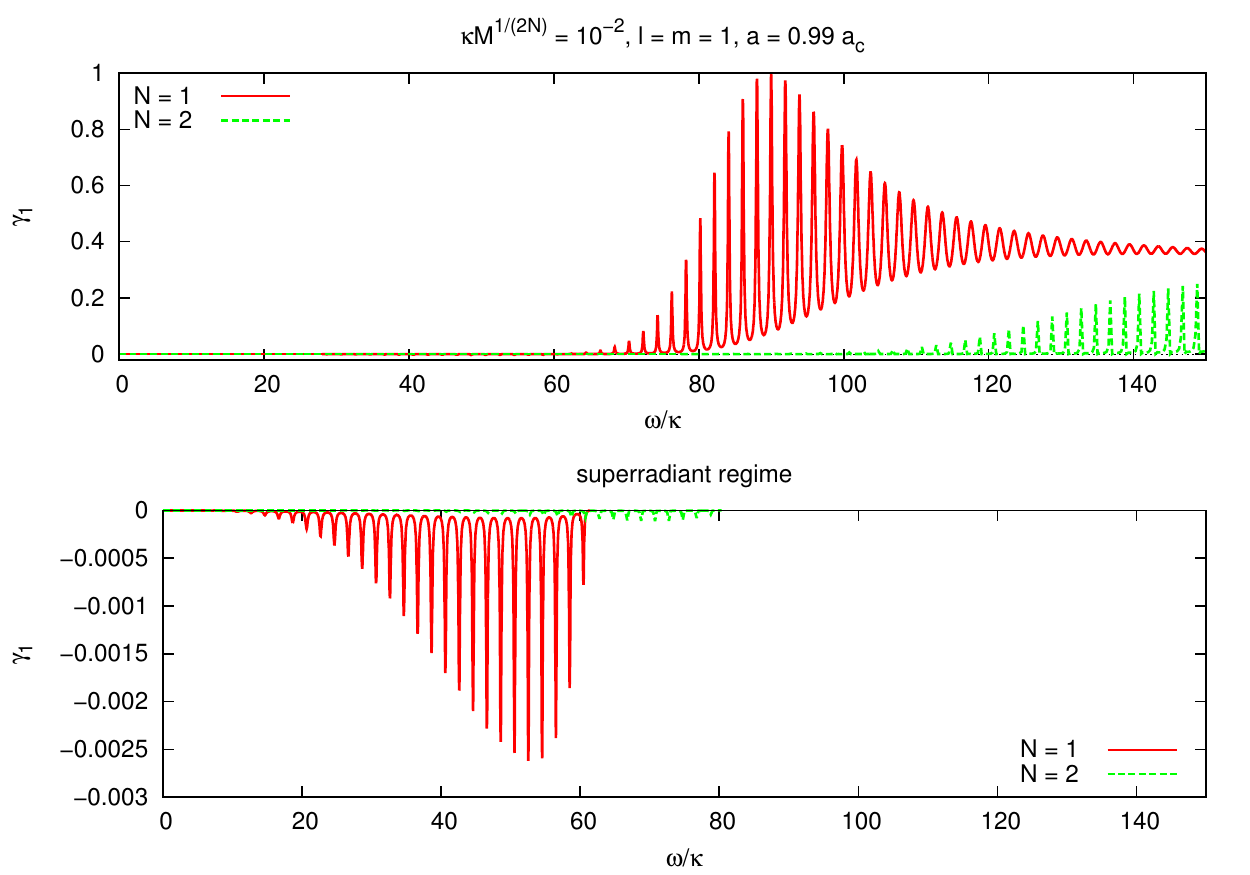}
\caption{\emph{Top panel:} Greybody factors (numerical) for asymptotically Anti--de Sitter black holes with $a = 0.99 a_c$, $\ell = m = 1$, $N = 1, 2, 3$, and $\kappa M^{1/(2N)} = 10^{-2}$. \emph{Bottom panel:} Zoom-in on the superradiant regime. As in the asymptotically flat and asymptotically de Sitter cases, the most effective superradiance occurs in the case $N = 1$. For the case shown, the maximum amplification factor is $0.26\,\%$.}
\label{fig:AdS_small-l1a0.99}
\end{figure}

\begin{figure}[H]
 \centering
 \includegraphics[width=11.5cm]{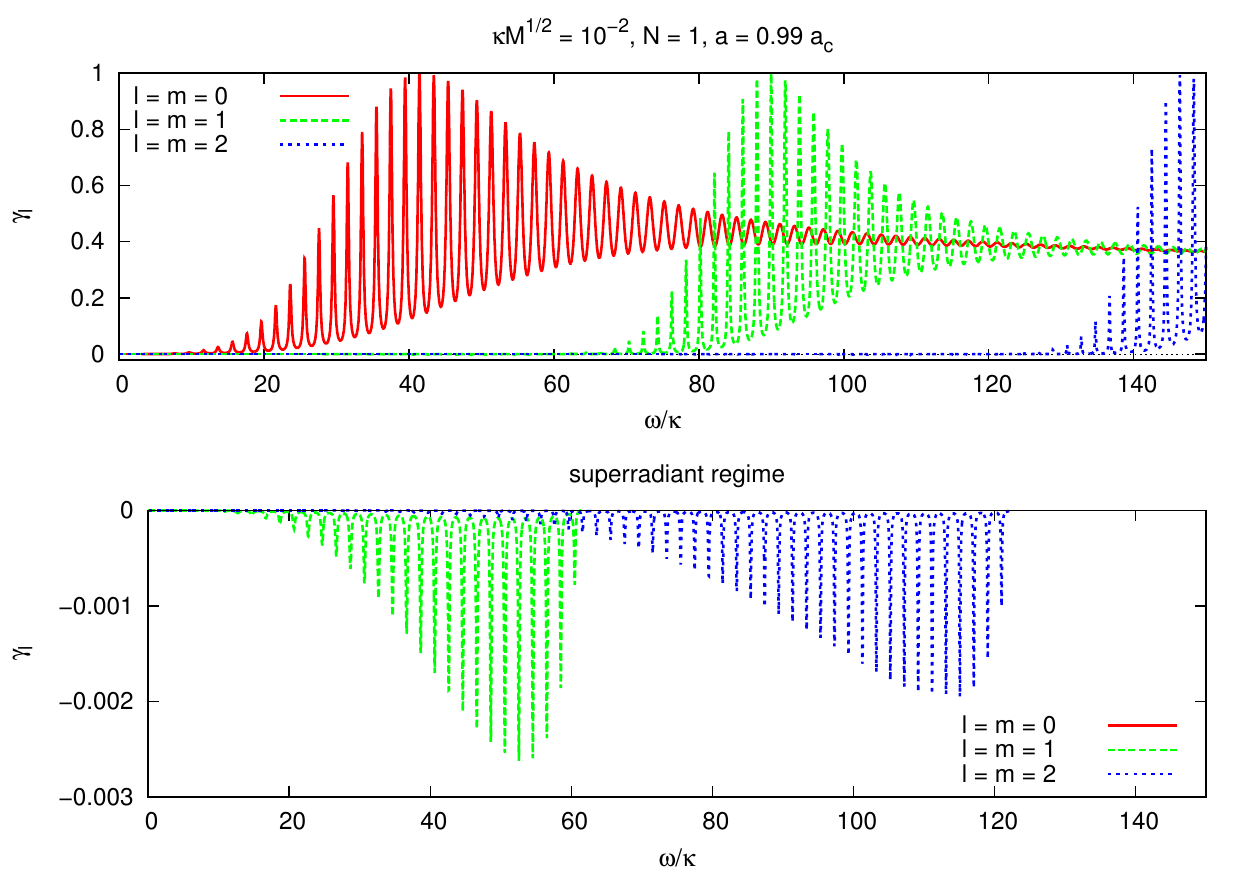}
 \caption{\emph{Top panel:} Greybody factors (numerical) for asymptotically Anti--de Sitter black holes with $N = 1$, $\kappa M^{1/2} = 10^{-2}$, $ \ell = m = 0, 1, 2$. \emph{Bottom panel:} Zoom-in on the superradiant regime. Superradiance is most effective in the case $ \ell = m = 1$.}
 \label{fig:AdS_small-N1a0.99}
\end{figure}

\begin{figure}[H]
 \centering
 \includegraphics[width=11cm]{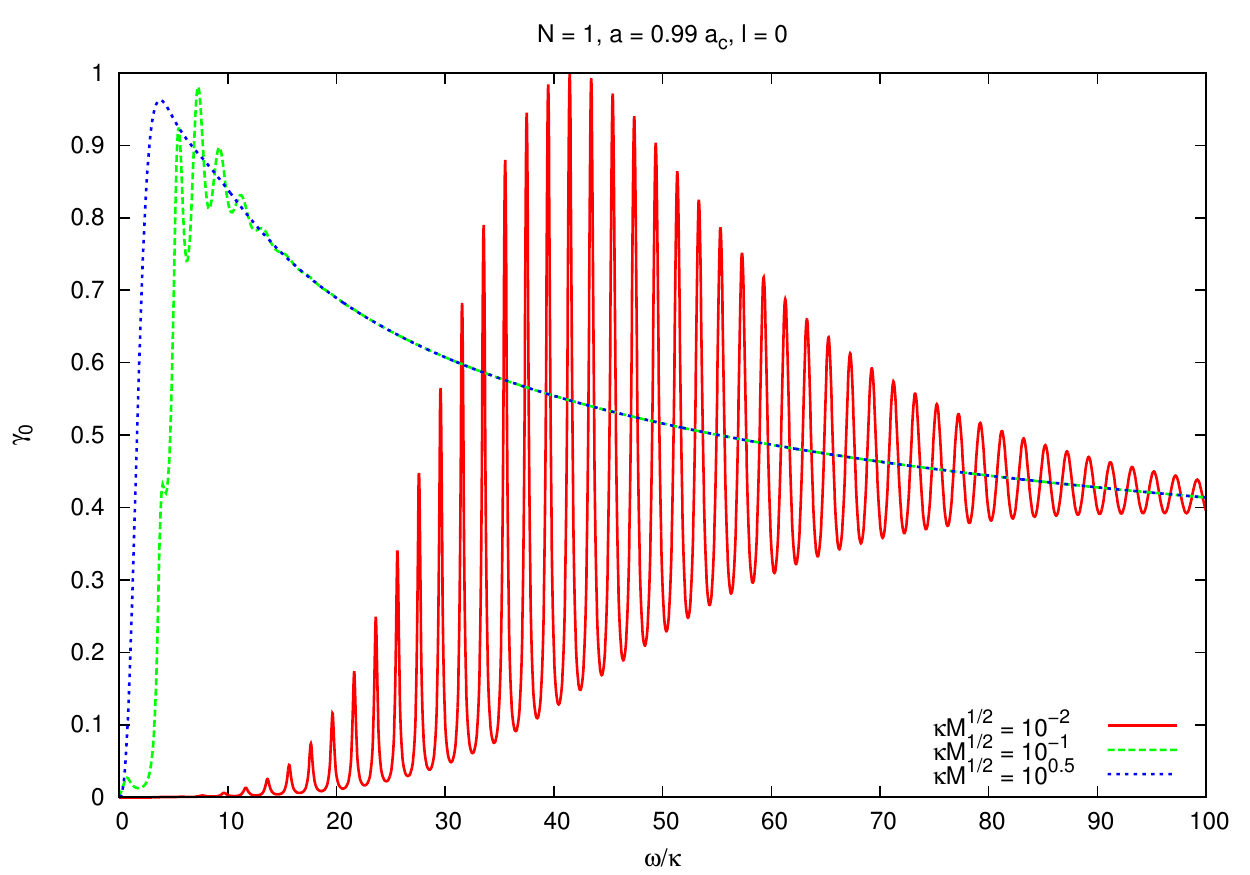}
 \caption{Greybody factors (numerical) for $\ell = 0$, $a = 0.99 \, a_c$, $N = 1$, and $\kappa M^{1/2} = 10^{-2}, 10^{-1},10^{0.5}$. The oscillations presumably related with pure AdS normal frequencies are prominent for small black holes but get damped as the black hole mass increases.}
 \label{fig:AdS_comp_gen}
\end{figure}

We present results for AdS black holes in Figs.~\ref{fig:AdS_comp}--\ref{fig:AdS_comp_gen}. Once again, we start by comparing numerical and analytic results for small black holes, Eq.~\eqref{eq:AdS_lf}, in Fig.~\ref{fig:AdS_comp}. We set $\kappa M^{1/(2N)} = 10^{-2}$, $\ell = 0$, and consider a variety of choices for $N$ and $a$. Excellent agreement is obtained between analytic and numerical results, independently of the choice of black hole configuration, i.e. parameters $\{N, M, a\}$, and this agreement improves as the frequency decreases. This is expected since the regime of validity of Eq.~\eqref{eq:AdS_lf} is in part determined by $\omega \ll T_H, r_h^{-1}$.

In Fig.~\ref{fig:AdS_small-l0a0.99} (left panel) we compare numerical results for $\kappa M^{1/2} = 10^{-3}$,  $N=1$ and $l = 0$, with varying choices of spin parameter $a$. It is clear that the greybody factor for the s-wave depends only weakly on the rotation. The right panel shows a similar comparison, but now fixing $\kappa M^{1/(2N)} = 10^{-2}$, $l = 0$ and $a = 0.99 \, a_c$, while varying the number of dimensions, $N = 1, 2, 3$. In the frequency interval shown, the greybody factor decreases with growing $N$. As discussed in Sec.~\ref{sec:smallAdS_analytic}, the peaks observed in the transmission coefficient are separated by intervals that are roughly consistent with the normal frequency spectrum of pure AdS, although the relation between the two is not clear. Indeed, the peaks apear slightly shifted relative to the normal frequencies of AdS [\textit{cf.} Eq.~\eqref{eq:AdSnormalfrequencies} and Fig.~\ref{fig:gbf_small_AdS_ana}]. Besides that, the amplitude of the peaks becomes larger as the frequency increases, whereas at low frequencies the presence of the black hole makes the greybody factors vanish in the zero-frequency limit.

Superradiant effects in small AdS black holes, $\kappa M^{1/(2N)} = 10^{-2}$, are evident in Fig.~\ref{fig:AdS_small-l1a0.99} for $\ell = m = 1$ and different dimensionalities, $N = 1, 2, 3$. As in the asymptotically flat and de Sitter cases, superradiance is suppressed as the number of dimensions increases. The propagation of the above-mentioned peaks to the superradiant regime can lead to large superradiant amplifications. For example, the maximum amplification factor obtained for $N = 1$ ($\kappa M^{1/2} = 10^{-2}$) is $0.26\,\%$, while for asymptotically flat black holes we obtained $0.046\,\%$, thus representing a five-fold enhancement.

In fact, significantly larger superradiant amplifications can be obtained if one continues to increase the size of the black hole. Typically, as one does so, the greybody factor grows more negative. This should be expected, assuming the universality for small black holes pointed out in Sec.~\ref{sec:smallAdS_analytic} is not significantly altered by the consideration of $\ell\neq0$ modes. This picture ceases to be accurate as soon as we leave the small black hole regime. Moreover, the superradiant frequency window shrinks as we increase the mass of the black hole, suggesting that superradiance is maximized for intermediate-sized black holes in AdS. Unfortunately, we have not been able to obtain high-resolution numerical results with fully satisfactory convergence for both this class and large black holes\footnote{Our numerical results for intermediate-size and large black holes in AdS show some dependence on the choice of numerical integration domain, especially in the superradiant frequency regime.}. Nevertheless, we do have indications that the maximum superradiant amplification, at least in $D=5$ dimensions, occurs for black holes with spin parameter very close to being maximal ($a=a_c$) and a mass (in units of the AdS scale) of order unity, more precisely for $\kappa^2 M\simeq 1.5$. Interestingly, Ref.~\cite{Cardoso:2013pza} also found that the equal-angular momenta AdS black holes in five dimensions which feature the strongest superradiant instabilities fall within this same ball park of parameters \footnote{We thank \'Oscar Dias for pointing out that the strongest superradiant instabilities occur for near-extremal --- as opposed to exactly extremal --- rotations.} (see Fig.~17 of Ref.~\cite{Cardoso:2013pza}). Moreover, our preliminary results suggest that for the intermediate-mass black holes in AdS the maximum amplification factor can reach values in excess of $100\,\%$. Clearly, this phenomenon deserves further dedicated study.

In Fig.~\ref{fig:AdS_small-N1a0.99} we show a comparison between modes with $\ell = m = 0, 1, 2$ for $N = 1$ and still for the class of small black holes in AdS, with $\kappa M^{1/2} = 10^{-2}$. As in all cases presented so far, superradiance is most effective for the $\ell = m = 1$ mode. We should also point out that results for $\ell = m = 0$ and $\ell = m = 1$ tend to the same value in the high-frequency regime, as expected on general grounds from the dependence of the effective potential~\eqref{eq:potential} on the parameters $\ell$ and $m$.

Results for $\ell = 0$, $N = 1$, $a = 0.99 \, a_c$, and black holes with different sizes ($\kappa M^{1/2} = 10^{-2}, 10^{-1},10^{0.5}$) are presented in Fig.~\ref{fig:AdS_comp_gen}. The local maxima in the greybody factor are numerous for small black holes, but they get diluted as the black hole mass increases. These peaks leave a strong imprint on the greybody spectrum, and they are a physical consequence of the coupling between the field and the negative cosmological constant. For high frequencies, greybody factors tend to zero in the same fashion, irrespective of the mass of the black hole. This is a consequence of the fact that the effective potential~\eqref{eq:potential} for AdS black holes diverges at infinity.

Finally, we would like to discuss how our results regarding superradiance in AdS black hole spacetimes fit within the general analysis of Ref.~\cite{Winstanley:2001nx}. We recall the conclusions of that study: superradiance is absent if (i) reflective boundary conditions are imposed, or (ii) transparent boundary conditions are used, the BH is rotating sufficiently slow (i.e., $\Omega_h< \ell^{-1}$), \emph{and a natural choice of ``positive" frequency modes is made}. This choice can only be made when a global timelike Killing vector exists, which is equivalent to $\Omega_h< \ell^{-1}$. In practice, the effect of making such a choice is to declare as ``positive" frequency modes only those frequencies for which $\omega-m\Omega_h > 0$. This factor is precisely the one that gives us the superradiant effect. If we disregarded it as not being part of the spectrum of positive frequencies we would observe no superradiance, in accordance with~\cite{Winstanley:2001nx}.

\section*{Acknowledgements}

It is a pleasure to thank Lu\'is Crispino for initial collaboration on this project and for numerous discussions. We thank Atsushi Higuchi and \'Oscar Dias for helpful correspondence.
J.~V.~R. also thanks Vitor Cardoso, Jos\'e Nat\'ario and Paolo Pani for useful discussions.
E.~S. de~O. is grateful to {\it Conselho Nacional de Desenvolvimento Cient\'\i fico e Tecnol\'ogico} (CNPq) and to {\it Coordena\c{c}\~ao de Aperfei\c{c}oamento de Pessoal de N\'\i vel Superior} (CAPES) for partial financial support.
J.~V.~R. is supported by {\it Funda\c{c}\~ao para a Ci\^encia e a Tecnologia} (FCT)-Portugal through Contract No.~SFRH/BPD/47332/2008.

\appendix
\section{Monotonicity of the horizon area\label{sec:Appendix}}

\setcounter{equation}{0}

In this appendix we prove that the horizon area for the cohomogeneity-1 black holes that we consider in this paper is a monotonically decreasing function of the rotation parameter $a$ (when considered as a function of the `bare' parameters $a$ and $M$).

We begin by computing the horizon area. From~\eqref{eq:metric}, the induced metric on the horizon is
\be
 ds_{hor}^2  \equiv {\mathfrak g}_{\mu\nu} dy^\mu dy^\nu =   r_h^2 \widehat{g}_{ab} dx^a dx^b +\, h(r_h)^2 \left[ d\psi + A_a dx^a \right]^2\,,
\qquad\qquad y^\mu=(\psi,x^a)\,.
\ee
The area of the horizon is thus, in $D=2N+3$ dimensions,
\be
 {\cal A}_h =  \int dx^\mu \sqrt{{\mathfrak g}} = r_h^{2N}h(r_h) \int_0^{2\pi}d\psi \int dx^a \sqrt{\widehat{g}}
 =  2\pi\, r_h^{2N}h(r_h) \, {\rm Vol}({\mathbb CP}^N) = \Omega_{2N+1}\, r_h^{2N}h(r_h) \,.
 \label{eq:area}
\ee
In the last equality we took into account that ${\rm Vol}({\mathbb CP}^N)=\pi^N/\Gamma(N+1)$.

Now we want to show that $\frac{d{\cal A}_h}{da}\leq0$\,. To this end observe that besides the explicit dependence on the spin parameter $a$ coming from the definition of function $h$, the horizon radius also has implicit dependence on $a$ since it is defined by
\be
g(r_h)^{-2}\equiv 1+\kappa^2r_h^2 - \frac{2M}{r_h^{2N}}\left(1-\kappa^2a^2-\frac{a^2}{r_h^2}\right) = 0\,.
\ee
By taking a derivative with respect to $a$ we obtain
\be
\frac{\partial r_h}{\partial a} = - \frac{2Ma}{r_h^{2N+1}} \frac{1+\kappa^2r_h^2}{\kappa^2r_h^2 + N \frac{2M}{r_h^{2N}}(1-\kappa^2a^2) - (N+1)\frac{2Ma^2}{r_h^{2N+2}}} \,.
\ee
Using this information and~\eqref{eq:area} we can evaluate the derivative $\frac{d{\cal A}_h}{da}$:
\beq
\frac{d{\cal A}_h}{da} &=& \Omega_{2N+1} \left\{ \frac{2Ma}{r_h^{2N+1}} + \left[ (2N+1) + N \frac{2Ma^2}{r_h^{2N+2}} \right]  \frac{\partial r_h}{\partial a} \right\} \nonumber\\
&=& -\, \Omega _{2N+1}\frac{2Ma}{h(r_h)} \, \frac{1+\frac{2M}{r_h^{2N}}\left(N+\frac{a^2}{r_h^2}\right)}{\kappa^2r_h^2 + N \frac{2M}{r_h^{2N}}(1-\kappa^2a^2) - (N+1)\frac{2Ma^2}{r_h^{2N+2}}} \,.
\label{eq:dAda}
\eeq
It is not obvious that the denominator is positive but this in fact follows from the requirement that the spacetime possesses a black hole horizon.\footnote{The extremality condition is attained when the horizon is degenerate, and so the derivative $\frac{d}{dr_h} [g(r_h)^{-2}]$ must vanish in the extremal case. But in general it must be non-negative,
\be
\frac{d}{dr_h} [g(r_h)^{-2}] = \kappa^2r_h^2 + N \frac{2M}{r_h^{2N}}(1-\kappa^2a^2) - (N+1)\frac{2Ma^2}{r_h^{2N+2}} \geq 0\,. \nonumber
\ee
}
Therefore, the right hand side of expression~\eqref{eq:dAda} is manifestly non-positive and we have proved that {\em the horizon area monotonically decreases with spin for this class of black holes in any odd dimension and with arbitrary cosmological constant}.

The statement above regarded the non-physical parameters $M$ and $a$ as independent variables characterizing the solution. The physical parameters are given in Eq.~\eqref{eq:charges}, which can be inverted to express the spin and mass parameters in terms of the mass and angular momentum of the spacetime. These black holes satisfy the first law of black hole thermodynamics~\cite{Gibbons:2004ai}
\be
d{\cal M} = T_H d{\cal S} +\Omega_h d{\cal J}\,,
\ee
where ${\cal S}$ stands for the entropy. Therefore we can easily conclude that, keeping the mass fixed, the area of the event horizon decreases with increasing angular momentum,
\be
\frac{d{\cal A}_h}{d{\cal J}} = 4G \frac{d{\cal S}}{d{\cal J}} = -4G \frac{\Omega_h}{T_H} \leq 0 \,,
\label{eq:dAdJ}
\ee
where we have reintroduced the gravitational constant $G$ for clarity.
This inequality can also be demonstrated by a more cumbersome evaluation of the derivative
\be
\left. \frac{d{\cal A}_h}{d{\cal J}}\right|_{\rm fixed\; {\cal M}} =  \frac{d{\cal A}_h}{da} \frac{da}{d{\cal J}} + \frac{d{\cal A}_h}{dM} \frac{dM}{d{\cal J}}
= \left( \frac{d{\cal A}_h}{da} - \frac{\kappa^2Ma}{N+\frac{1}{2}+\frac{\kappa^2a^2}{2}} \frac{d{\cal A}_h}{dM} \right) \frac{da}{d{\cal J}} \,.
\ee
The derivative $\frac{d{\cal A}_h}{dM}$ can be computed just like we did for $\frac{d{\cal A}_h}{da}$. It turns out that the combination within brackets is non-positive while $\frac{da}{d{\cal J}}$ is positive. Therefore we reproduce the result~\eqref{eq:dAdJ}.

\bigskip


\end{document}